\documentclass [a4paper, 12pt]{article}
\usepackage[utf8]{inputenc}
\usepackage{booktabs}
\usepackage{xcolor}
\definecolor{darkred}{rgb}{0.4,0.0,0.0}
\definecolor{darkgreen}{rgb}{0.0,0.4,0.0}
\definecolor{darkblue}{rgb}{0.0,0.0,0.4}
\usepackage[bookmarks,linktocpage,colorlinks,
    linkcolor = darkred,
    urlcolor  = darkblue,
    citecolor = darkgreen]{hyperref}

\usepackage{subfigure}
\usepackage{amssymb}
\usepackage{amsmath}
\usepackage{epsfig}
\usepackage{graphicx}
\usepackage{times}
\usepackage{float}
\usepackage{hyperref}
\usepackage{multirow}
\usepackage{slashed}
\usepackage{enumitem}
\usepackage{bbold}
\usepackage[small]{caption}
\usepackage{fullpage}
\usepackage{cite}
\usepackage{algorithm}
\usepackage{algorithmic}
\numberwithin{equation}{section}

\newcommand{\norm}[1]{\left\lVert#1\right\rVert}
\newcommand{\abs}[1]{\left|#1\right|}
\newcommand{\PCAC}{\text{\scriptsize{PCAC}}}
\newcommand{\ND}{\text{\scriptsize{ND}}}
\newcommand{\TM}{\text{\scriptsize{TM}}}

\newcommand{\W}{\text{\scriptsize{W}}}

\newcommand{\cn}{\text{cn}}
\newcommand{\sn}{\text{sn}}
\newcommand{\cs}{\text{cs}}

\title{Multigrid approach in shifted linear systems for the non-degenerated twisted mass operator
}
\author{Constantia Alexandrou\textsuperscript{a,b}, Simone Bacchio\textsuperscript{a,c}, 
Jacob Finkenrath\textsuperscript{b}\\[0.2cm]
\small\textsuperscript{a}Department of Physics, University of Cyprus, PO Box 20537, 1678 Nicosia, Cyprus\\
\small\textsuperscript{b}Computation-based Science and Technology Research Center, The Cyprus Institute\\ 
\small\textsuperscript{c}Fakult\"at f\"ur Mathematik und Naturwissenschaften, Bergische Universit\"at Wuppertal}
\date{}

\begin{document}

\maketitle

\begin{abstract}
Application of multigrid solvers in shifted linear systems is studied.
We focus on accelerating the rational approximation needed for simulating single flavor operators. 
This is particularly useful, in the case of twisted mass fermions for mass non-degenerate quarks and can be employed  
to accelerate the $N_f=1+1$ sector of $N_f=2+1+1$ twisted mass fermion simulations.
The multigrid solver is accelerated by employing suitable initial guesses.
We propose a novel strategy for proposing initial guesses for shifted linear
systems based on the Lagrangian interpolation of the previous solutions.
\end{abstract}

\section{Introduction}\label{intro}

Simulation of lattice Quantum Chromodynamics (QCD) is computationally very  demanding 
due to the requirement of solving a large number   of linear system equations, 
which involve very large sparse complex matrices.
A specific kind of these linear equations of focus in this paper, is
given by multi-shifted systems
\begin{equation} 
 (Q^2 + m_i^2I)\,x_i = b
\label{eq:shifted_linear}
\end{equation}
for a set of different shifts $m_i$ acting on the diagonal of the squared operator $Q$ and where all shifts
depend on a common right hand side $b$.
Speeding-up the time to solution is essential for performing the simulation
and we discuss how this is achieved within 
a multigrid approach. Multigrid approaches
can be efficient for the case that the operator $Q$ develops very small eigenvalues, thus
yielding iteration counts of $\mathcal{O}(1000)$ when using approaches based on
non-preconditioned Krylov subspace methods. 

Shifted linear systems, here referred to as multi-mass shifted systems,
have to be solved if the square root of a matrix is evaluated via rational, polynomial
approximations (see e.g.~\cite{Frezzotti:1997ym,Clark:2006fx}) or by using an integral definition via Stieltjes function~\cite{Frommer2014a,Frommer2014b}.
In general the matrix roots have to be calculated in case of Monte Carlo simulations involving a single 
quark~\cite{Clark:2006fx}. Another application of the matrix square root is given by the calculation
of the Neuberger overlap operator \cite{Neuberger:1997fp}.

The approaches routinely used to solve shifted linear equations like Eq.~\eqref{eq:shifted_linear}
are given by Krylov space solvers ~\cite{Jegerlehner:1996pm,Glassner:1996gz, Frommer:1998,Frommer2014a,Frommer2014b}.
In what follows we will refer to the multi-mass shift conjugated gradient (MMS-CG) solver as the ``standard'' solver \cite{Jegerlehner:1996pm,Glassner:1996gz}.
The advantage of multi-mass shifts solvers is that the numerical effort scales only
with the cost for solving the smallest shift and thus is almost independent of the total number of shifts.
The idea behind these approaches is to exploit the identical eigenspaces
by using the same Krylov space generated for the most ill-conditioned
system to iterate all other shifts. However, if the operator $Q^2$ develops very small eigenvalues, 
flexible preconditioned iterative solvers can be by far more effective as compared to the conjugate gradient solver.

The central idea of this paper is to introduce a hybrid method, using 
MMS-CG solver for systems with a larger mass-shift,
while for the systems with the smaller shifts employ a multigrid approach using extrapolated initial guesses.
The novelty of this method is how these initial
guesses are constructed. We will discuss two approaches, one using Lagrange interpolation
based on the solution of other shifts and the second using the MMS-CG solver to construct
guesses which will be used for the systems involving the smallest mass-shifts.

For very ill-conditioned systems
a very effective approach is given by multigrid methods as a preconditioner for Krylov solvers~\cite{Luscher:2007se,Babich:2010qb, Frommer:2013fsa}.
Multigrid solvers optimized for  lattice QCD fermion operators show a very good scaling
down to the physical light quark mass and can speed-up the time to solution by more than two order of
magnitude compared to standard Krylov methods like the conjugate gradient (CG) solver.
The idea is to use a coarse grid correction to tackle small eigenvalues and a smoother to
tackle large eigenvalues in each iteration.
However, all multigrid approaches are  optimized to solve
a linear equation where only a single operator is involved, as given by $Q x= b$.
This will not hinder the solution of the squared systems but introduces different
ways to tackle the multi-mass shifted system of Eq.~\eqref{eq:shifted_linear}
given by two consecutive solves of the single systems or taken the difference of two solve of the single system.
Moreover, given the fact that state-of-the-art simulations are being performed using increasingly larger lattice sizes, 
a closer look to the different ways by discussing the involved stopping criteria
seems advisable.

Multigrid approaches in lattice QCD are known for a variety of fermion discretizations,
such as for the Wilson fermion discretization~\cite{Luscher:2007se,Babich:2010qb, Frommer:2013fsa}, as a preconditioner 
for Neuberger overlap fermions~\cite{Brannick:2014vda}, for Domain Wall fermions \cite{Cohen:2012sh}
or for staggered fermions~\cite{Brower:2018ymy}.
In this work we consider  twisted mass fermions~\cite{Frezzotti:2003xj} and
extend the approach of Ref.~\cite{Alexandrou:2016izb}
for the more general case of using a doublet of twisted mass fermions with different masses.
As in Ref.~\cite{Alexandrou:2016izb}, we adapt the adaptive aggregation-based domain decomposition multigrid (DD-$\alpha$AMG) approach
and show how the symmetries of the non-degenerated twisted mass operator can be successfully
used to build a coarse grid correction.

The paper is organized as follows:
In section~\ref{sec:1p1TM} we  discuss how the fermion operator, here represented by $Q$,
can be constructed on a four dimensional lattice, giving  in detail the representation
of a doublet of twisted mass fermions with different masses.
In section~\ref{sec:DDalphaAMG} we explain the DD-$\alpha$AMG approach 
and how it can be used to solve a linear equation that involves a squared
operator $Q^2 x = b$,  especially when  large lattice sizes are used and
in the presence of small 
quark masses.
We then present how the DD-$\alpha$AMG approach can be adapted
to the non-degenerated (ND) twisted mass operator.
This involves a discussion of the properties of the ND twisted mass operator
and how this can be used to construct a coarse grid corrections. At the end of
the section we  show  numerical results.
In section~\ref{sec:MGinit_guess} we present the central idea of this work, namely the solution of
the multi-mass shifted systems via a hybrid approach. For that we first 
introduce the approximation used for the square root matrix function, given
by a rational approximation and explain
how to construct initial guesses via Lagrange interpolation and from the MMS-CG solver. 
To illustrate the effectiveness of our approach
we present results using $N_f=2+1+1$  gauge-configurations
at physical quark masses.

\section{Wilson twisted mass fermions}\label{sec:1p1TM}

Solving  Eq.~\eqref{eq:shifted_linear}, requires the inverse
of a very large sparse matrix, which depends on the  discretization scheme of the fermionic action.
We consider here the two flavor non-degenerated twisted mass operator given by~\cite{Frezzotti:2003xj}
\begin{eqnarray}
D_{\ND}(\bar\mu,\bar\epsilon) = (D_\W\otimes I_2)+i\bar\mu\,(\Gamma_5\otimes\tau_3) -\bar\epsilon ( I \otimes \tau_1) &=&
\begin{bmatrix}
D_\W+i\bar\mu\Gamma_5        & -\bar\epsilon I \\
-\bar\epsilon I & D_\W-i\bar\mu\Gamma_5
\end{bmatrix} \notag \\
&=&
\begin{bmatrix} D_{\TM}(\bar\mu) & -\bar\epsilon I \\
-\bar\epsilon I  & D_{\TM}(-\bar\mu)  \\
\end{bmatrix}\label{eq:non-degenerate}
\end{eqnarray}
with the bare mass parameters $\bar\mu$ and $\bar\epsilon \in \mathbb{R}$.
The twisted mass terms $\bar\mu$ and $\bar\epsilon$ act on the two-dimensional flavor space and $\tau$ are the Pauli matrices.
\begin{equation}
\tau_1 = \begin{pmatrix}
\phantom{1} & 1 \\
1  & 
\end{pmatrix}
\quad\text{and}\quad
\tau_3 = \begin{pmatrix}
1 & \phantom{1} \\
& -1
\end{pmatrix}.
\end{equation}
The twisted mass (TM) Wilson Dirac operator
\begin{equation}\label{eq:D(mu)}
D_{\TM}(\mu) = D_\W +i \mu \Gamma_5
\end{equation}
acts on the single flavor space  $\mathcal{V}_s = \mathcal{V} \times\mathcal{S}\times\mathcal{C}$
with $\mathcal{S}$ the spin space and $\mathcal{C}$ the color space.
The spatial volume $\mathcal{V}=T\cdot L^3$ is the four-dimension hyper-cubic lattice defined by
\begin{equation}
\mathcal{V} = \{ x = (x_0,x_1,x_2,x_3), 1 \leq x_0 \leq T, \, 1 \leq x_1,x_2,x_3 \leq L \}
\end{equation}
with $T$ the number of points in the temporal direction and 
$L$ the number of points in the spatial directions $x,\,y$ and $z$.
The matrix $\Gamma_5 = I_\mathcal{V}\otimes\gamma_5\otimes I_\mathcal{C}$ is based on the spin space component $\gamma_5$, 
which can be represented by $\gamma_5 =  \gamma_0 \gamma_1 \gamma_2 \gamma_3 = diag\{ \phantom{-}1 , \phantom{-}1, -1, -1\}$
in the so-called chiral representation.
Note that we set the lattice spacing $a$ to unity throughout this paper.
The TM Wilson operator is itself based on the Wilson Dirac operator $D_\W \equiv D_\W(U,m,c_{sw})$
given by
\begin{align}\label{eq:Wilson}
	(D_\W\psi) (x) =& \Big( (m+4) I_{12} - \frac{c_{sw}}{32} \sum_{\mu,\nu=0}^3 ( \gamma_\mu \gamma_\nu ) \otimes \big( Q_{\mu\nu}(x) - Q_{\nu\mu}(x) \big) \Big) \psi(x) \nonumber\\
	&  \mbox{} - \frac{1}{2}\sum_{\mu=0}^3 \left( (I_4-\gamma_\mu)\otimes U_\mu(x)\right) \psi(x+\hat{\mu}) \nonumber\\
	& \mbox{} - \frac{1}{2}\sum_{\mu=0}^3 \left( (I_4+\gamma_\mu)\otimes U_\mu^\dagger(x-\hat{\mu})\right) \psi(x-\hat{\mu}) \;,
\end{align}
with $m$ the mass parameter and $c_{sw}$ the parameter of the clover term~\cite{Sheikholeslami:1985ij}, 
~for a definition of $Q_{\nu\mu}(x)$ see e.g.~Ref.~\cite{Alexandrou:2016izb}.
The gauge links $U_\mu(x)$ are $SU(3)$ matrices,
and the set $\{ U_\mu(x) \, : \, x \in \mathcal{L} , \, \mu=0,1,2,3 \}$ is referred to as a gauge configuration.
The $\gamma$-matrices act on the spin degrees of freedom of the spinor field $\psi(x)$
and fulfill the anti-commutation relation $\{\gamma_\mu,\gamma_\nu\} = 2 \cdot I_4 \; \delta_{\mu \nu}$.

Often the Wilson Dirac operator is decomposed using an even-odd reduction.
This is possible because the operator couples only  next neighboring points. Even-odd reduction has several advantages
compare to the full operator: it reduces the dimension of the operator and increases effectively the smallest eigenvalue
by a factor two (see e.g.~\cite{Alexandrou:2016izb}). Even-odd reduction is commonly used in  lattice QCD 
and in what follows we will use it throughout
for the ND twisted mass operator. This yields an overall
speed up for the time to solution in the case of the CG solver by a factor two. However, if a multigrid solver
is used the linear equation system with the full operator is solved
because the complexity of the structure increases in case of the even-odd reduced operator
due to next-to-next neighbor interaction. For more details we refer to Ref.~ \cite{Alexandrou:2016izb}.

In this paper, we consider the ND twisted mass operator for $\bar\epsilon>0$.
Due to the flavor mixing, introduced by the off-diagonal term proportional to $\bar\epsilon$, 
the renormalized quark masses for the strange and charm doublet, $m_s$ and  $m_c$ respectively, are connected to the bare twisted mass parameters by
\begin{equation}
m_s = \frac{1}{Z_P}\mu_s = \frac{1}{Z_P} \bar{\mu} - \frac{1}{Z_S} \bar{\epsilon} \qquad \textrm{and} \qquad m_c =  
\frac{1}{Z_P} \mu_c = \frac{1}{Z_P} \bar{\mu} + \frac{1}{Z_S} \bar{\epsilon}
\end{equation}
where $Z_S$ is the scalar and $Z_P$ is the pseudoscalar renormalization constants. 
For $\bar{\epsilon} =0$ the masses are degenerated, as currently employed in the simulations  for the light quark sector
with a mass-degenerated quark doublet representing the up- and down-quark.
The twisted mass discretization is free of ${\cal O} (a)$ effects if the light quark sector
is tuned to maximal twist~\cite{Frezzotti:2003ni,Frezzotti:2004wz}.  This is achieved by
tuning the Partially Conserved Axial Current (PCAC) mass, $m_{\PCAC}$,
to zero, i.e by demanding that  $m_{\PCAC} \rightarrow 0$.

\begin{figure}
	\centering
	\includegraphics[width=0.45\textwidth]{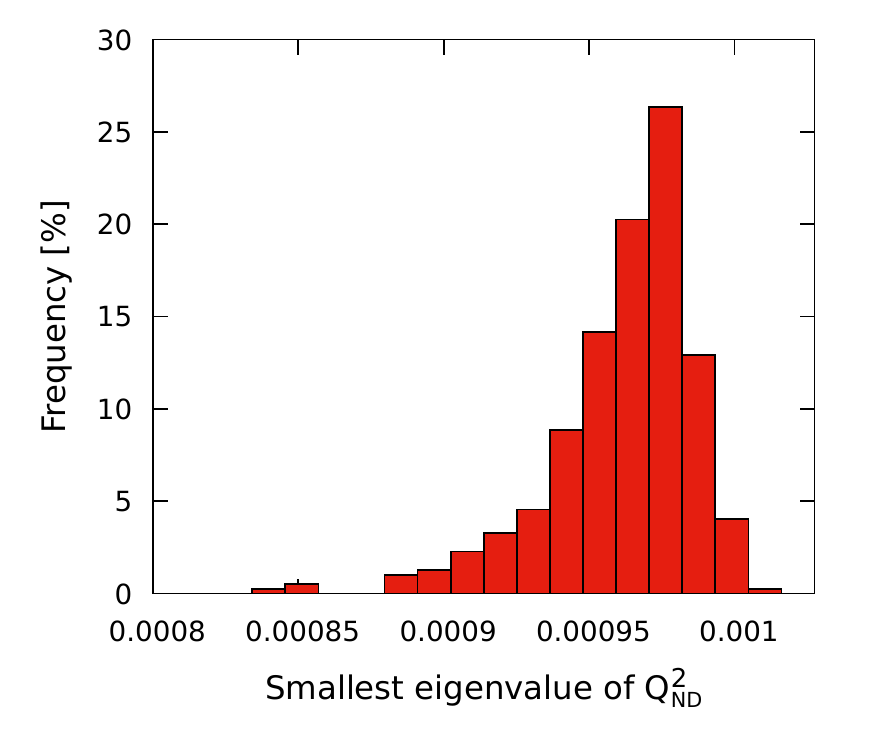}~~
	\includegraphics[width=0.45\textwidth]{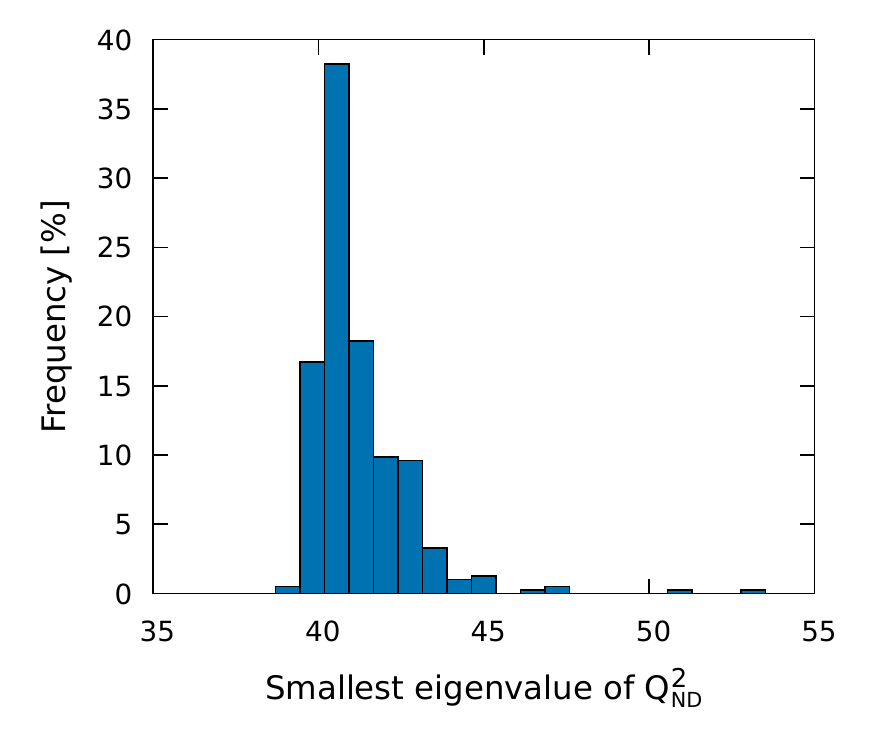}
	\caption{Distribution of the smallest (left) and largest (right) eigenvalues of the squared even-odd reduced
	        ND twisted mass operator using an ensemble of gauge-configurations tuned to the physical values of the light, strange and charm quark masses.
		}
	\label{fig:1+1_eig}
\end{figure}

The ND twisted mass operator is $(\Gamma_5\otimes \tau_1)$-hermitian:
\begin{equation}\label{eq:g5tau1-hermiticity}
Q_{\ND}^{\phantom\dagger}(\bar\mu,\bar\epsilon) = (\Gamma_5\otimes \tau_1)\,D_{\ND}^{\phantom\dagger}(\bar\mu,\bar\epsilon) = D_{\ND}^\dagger(\bar\mu,\bar\epsilon)\,(\Gamma_5\otimes \tau_1)= Q_{\ND}^\dagger(\bar\mu,\bar\epsilon).
\end{equation}
due to the $\Gamma_5$-hermiticity of the Wilson Dirac operator
\begin{equation}\label{eq:g5-hermiticity}
Q_\W = \Gamma_5 D_\W = D_\W^\dagger \Gamma_5 = Q_\W^\dagger.
\end{equation}
For the determinant of the ND twisted mass operator, it follows that 
\begin{equation}
\det\left[D_{\ND}(\bar\mu,\bar\epsilon)\right] \in \mathbb{R}
\end{equation}
using the  $(\Gamma_5\otimes \tau_1)$-hermiticity.
This implies the positiveness of the determinant.
The positiveness of the operator~\cite{Frezzotti:2003ni} is achieved for 
$\bar{\mu} > \bar{\epsilon}$. In case of the heavy quark doublet,
for physical values of the strange and charm quark masses, this is achieved if the ratio of the renormalization
constants is given by $Z_P/Z_S > 0.85$. 
However, numerically it is found that this bound is too strict, e.g.~
we found a gap for the smallest eigenvalue for the case of $Z_P/Z_S = 0.8$
as shown in Fig.~\ref{fig:1+1_eig}. We measured the distribution of the 
smallest and largest eigenvalues of the squared even-odd reduced ND twisted mass operator
on an ensemble of gauge configuration with a lattice volume of $V=128\times 64^3$
and a finite lattice spacing of $a\sim 0.08\; \textrm{fm}$, discussed in detail in \cite{Alexandrou:2018}.
We will denote this ensemble as our physical test ensemble and it will be 
used throughout this paper to numerically validate
our approach. It is generated with fermion parameters, such that
the light, strange and charm quark masses are tuned close to their physical values,
with $\mu=0.00072$ being the light quark mass parameter
and  with $\bar\mu=0.12469$ and $\bar \epsilon = 0.13151$ being
the heavy quark bare mass parameters of the ND twisted mass operator. 
Moreover in the tests we employ the ND twisted mass operator also for the light quark sector
using $\bar \mu_\ell = 0.00072$ and $\bar \epsilon_{\ell} = 0.000348$. 

\section{DD-$\alpha$AMG approach for twisted mass Wilson fermions}\label{sec:DDalphaAMG}

There exist several multigrid approaches for the Wilson Dirac operator~\cite{Luscher:2007se,Babich:2010qb, Frommer:2013fsa}.
In this section, we  outline the adaptive aggregation-based domain decomposition multigrid (DD-$\alpha$AMG) method
for the Wilson Dirac operator~\cite{Frommer:2013fsa}
and TM operator~\cite{Alexandrou:2016izb}.
The method is based on a flexible iterative Krylov solver, 
which is preconditioned at every iteration step by a multigrid approach.
The preconditioner acts on the error propagation through
\begin{equation}\label{eq:error_propagation}
\epsilon\,\leftarrow\,\left(I-MD_\W^{\phantom{1}}\right)^k\left(I-PD_{\W,c}^{-1}P^\dagger D_\W^{\phantom{1}}\right)\left(I-MD_\W^{\phantom{1}}\right)^j\epsilon,
\end{equation}
where the coarse grid correction $PD_{\W,c}^{-1}P^\dagger$ is applied
with $k$ pre- and $j$ post-iteration of the smoother $M$.
The multigrid preconditioner exploits domain decomposition 
strategies: the smoother $M$ is given by the Schwarz Alternating Procedure (SAP)~\cite{Luscher:2003qa}
and the coarse grid operator $D_{\W,c} = P^\dagger D_\W P$ is based on an aggregation-based projection
with the prolongation operator $P$. 
The method is designed to deal efficiently with both, infrared (IR) and ultra-violet (UV) modes of the operator $D_\W$. 
The smoother reduces the error components belonging to the UV-modes~\cite{Frommer:2013fsa}, while the
coarse grid correction is designed to deal with the IR-modes.
This is achieved by using an interpolation operator $P$, which approximately 
spans the eigenspace of the small eigenvalues. 
Thanks to the property of local coherence~\cite{Luscher:2007se}, 
the subspace can be approximated by aggregating over a small set of $N_v\simeq\mathcal{O}(20)$ test vectors $v_i$, which are computed in the DD-$\alpha$AMG method via an adaptive setup phase~\cite{Frommer:2013fsa}.
We remark that the prolongation operator  is $\Gamma_5$-compatible, i.e.~$\Gamma_5P = P\Gamma_{5,c}$.
This guarantees the $\Gamma_5$-hermiticity of the coarse grid operator $D_c$,
i.e.~$D_c^\dagger = \Gamma_{5,c}D_c \Gamma_{5,c}$ with $\Gamma_{5,c} = I_{\mathcal{V}_c}\otimes\sigma_3\otimes I_{N_v}$ being
the coarse grid equivalent of $\Gamma_5$.
On the coarsest grid, even-odd reduction
is used to accelerate the time solution.

The DD-$\alpha$AMG approach has been adapted to the TM Wilson
operator $D_{\TM}(\pm\mu)$ by modifying  the coarse grid procedure 
in order to avoid the slowing down of the time to solution for small $\mu\ll1$ by the coarse grid correction step. 
This is circumvent by effectively reducing the iterations for the coarse grid 
by using a larger $\mu$ for the coarse operator. i.e.~$D_{\TM,c}(\pm d\mu) =  D_{\W,c} \pm i d\mu\Gamma_{5,c}$ with $d \geq 1$~\cite{Alexandrou:2016izb}.
For $d\sim 5$ the iterations count for inverting the coarsest operator is reduced by an order
of magnitude while it affects only marginally the outer solver iterations count. 
This yields a compatible speed-up for the time to solution when using the DD-$\alpha$AMG approach for TM Wilson fermions as was obtained for the case of Wilson fermions~\cite{Frommer:2013fsa}.

\subsection{Solving the squared linear equation}
\label{sec:sqsol}

\begin{figure}
	\centering
	\includegraphics[width=0.7\textwidth]{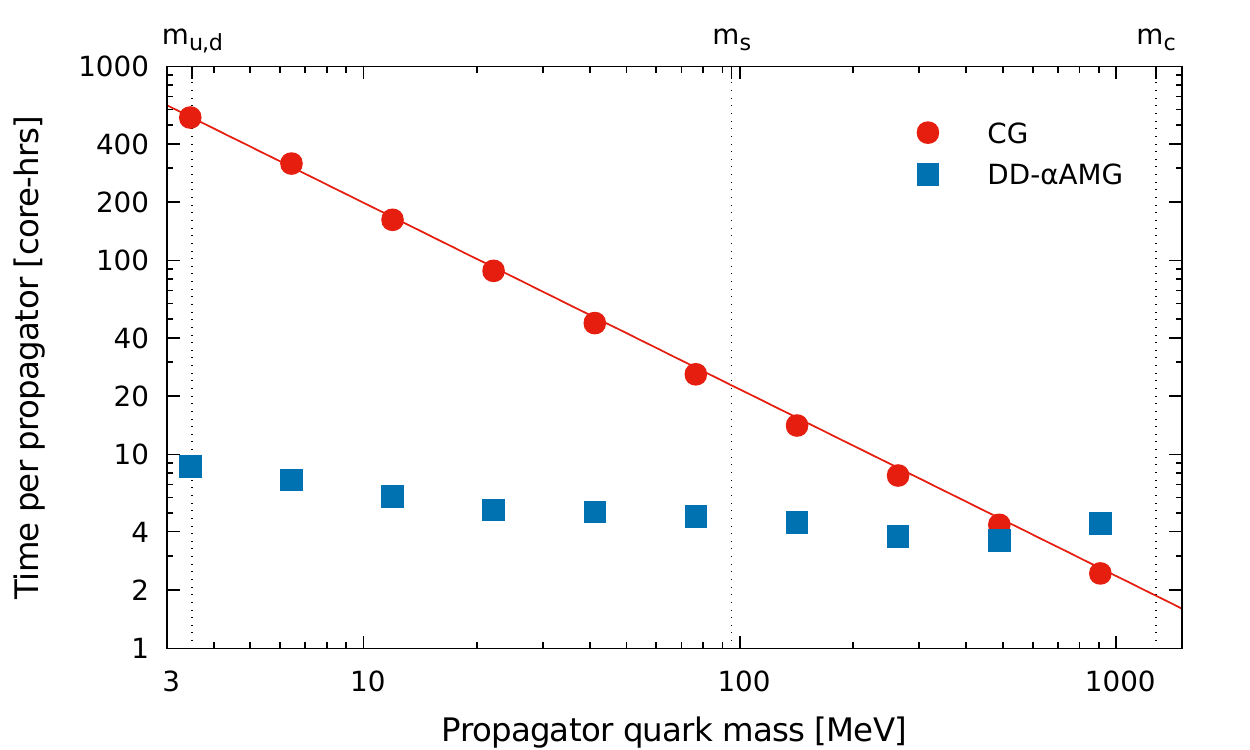}
	\caption{Comparison between time to solution for computing the twisted mass 
	fermion propagator at different quark mass using the odd-even (oe) CG solver and the DD-$\alpha$AMG approach. The results are
	for a $128\times64^3$ ensemble simulated at physical pion mass, $m_\pi\sim 135$~MeV. 
	The value of the light, strange and charm quark masses are shown by the
        vertical lines with the labels $m_{u,d}$, $m_s$ and $m_c$, respectively.}
	\label{fig:crit_slow_down}
\end{figure}

Before we introduce the new, adapted, multigrid approach for the ND twisted mass operator,
we  make some general remarks about solving a linear equation  involving a squared Dirac operator.
Solving the linear Dirac equation
involving a hermitian operator is needed within lattice QCD in several places such as for 
the fermion force calculation during the Hybrid Monte Carlo (HMC) simulation or 
the calculation of matrix functions, like the square root
or the sign function of the Dirac operator.
Let us consider linear equations of the form
\begin{equation}
 (Q_\W^2 + \mu^2)  x = b
 \label{eq:sqlineq}
\end{equation}
involving a squared operator $Q_\W^2$, which in our case, 
is the square of the TM Wilson operator. We will denoted
here the direct solution of eq.~\eqref{eq:sqlineq} as case \textbf{A}.
A standard approach to solve such equation is to use the conjugate
gradient (CG) solver. However, when one approaches the 
physical value of the light quark mass, the efficiency of the CG solver suffers
due to the increase of the number of small eigenvalues. In this parameter range, multigrid approaches are by far
more effective as shown in Fig.~\ref{fig:crit_slow_down} for the  case of the TM Wilson operator.
However, the multigrid approach outlined in the previous section, can only be applied
to a linear equation with a single TM Wilson operator. The reason is simply that an implementation
for the squared operator is far more complicated, involving next to next neighbor interactions,
which increase the complexity of the coarse operator.
A way to solve  Eq.~\eqref{eq:sqlineq}
is to modify it by exploiting the $\Gamma_5$-hermiticity
resulting into two sequential inversions, denoted here as case \textbf{B},
\begin{equation}
(Q_{\W}^2+\mu^2)^{-1} b  = (Q_{\W}\pm i\mu)^{-1} (Q_{\W}\mp i\mu)^{-1} b~.
\label{eq:linmul}
\end{equation}
Note that this can be done
in a similar way for the ND operator using the $(\Gamma_5\otimes\tau_1)$-hermiticity.
However, splitting up Eq.~\eqref{eq:sqlineq} is not unique 
and another possibility is given by a difference of the two single inversions,
denoted here as case \textbf{C},
\begin{equation}
\label{eq:lindif}
(Q_{\W}^2+\mu^2)^{-1} b  = \frac{i}{2\mu}\left((Q_{\W}+ i\mu)^{-1} b-(Q_{\W}- i\mu)^{-1} b\right)~.
\end{equation}
In exact arithmetic all ways yield to the same solution.
This is not so obvious anymore if iterative solvers are used since they involve 
a stopping criterium. A standard procedure is
to stop the solver if the relative residual $\norm{r}/\norm{b}$ is smaller than a chosen bound
$\rho$. The residual is given by  $r= Q x - b$
with $Q$ the involved operator and $b$ the involved rhs.

Analytically comparing the bounds of the residual of the iterated solution of Eq.~\eqref{eq:linmul} and Eq.~\eqref{eq:lindif} 
one would be in favor of case \textbf{B} for $\mu<1$. Indeed following Appendix \ref{sec:error} for 
the residuals we obtain
\begin{equation}
\frac{\norm{r_B}}{\norm{b}} \le \left(1+\sqrt{ \frac{\lambda^2_{\max} + \mu^2}{\lambda^2_{\min} + \mu^2}}\right)\rho  \qquad \textrm{and} \qquad \frac{\norm{r_C}}{\norm{b}} \le  \frac{\rho\sqrt{\lambda^2_{\max} + \mu^2}}{\abs{\mu}}~,  
\end{equation}
where each inversion is stopped when the relative residual fulfills $r/b<\rho$.
However, comparing the error
\begin{equation}
 \frac{\norm{e_B}}{\norm{b}} \le \frac{\rho}{{\lambda^2_{\min} + \mu^2}}  \qquad \textrm{and} \qquad \frac{\norm{e_C}}{\norm{b}} \le \frac{\rho}{\abs{\mu}\sqrt{\lambda^2_{\min} + \mu^2}}.
\end{equation}
it follows the error $e_B$ of case B is numerical equivalent to the error $e_C$ of case C
if $\lambda_{min} \ll \mu$.

\begin{figure}
	\centering
	\includegraphics[width=0.7\textwidth]{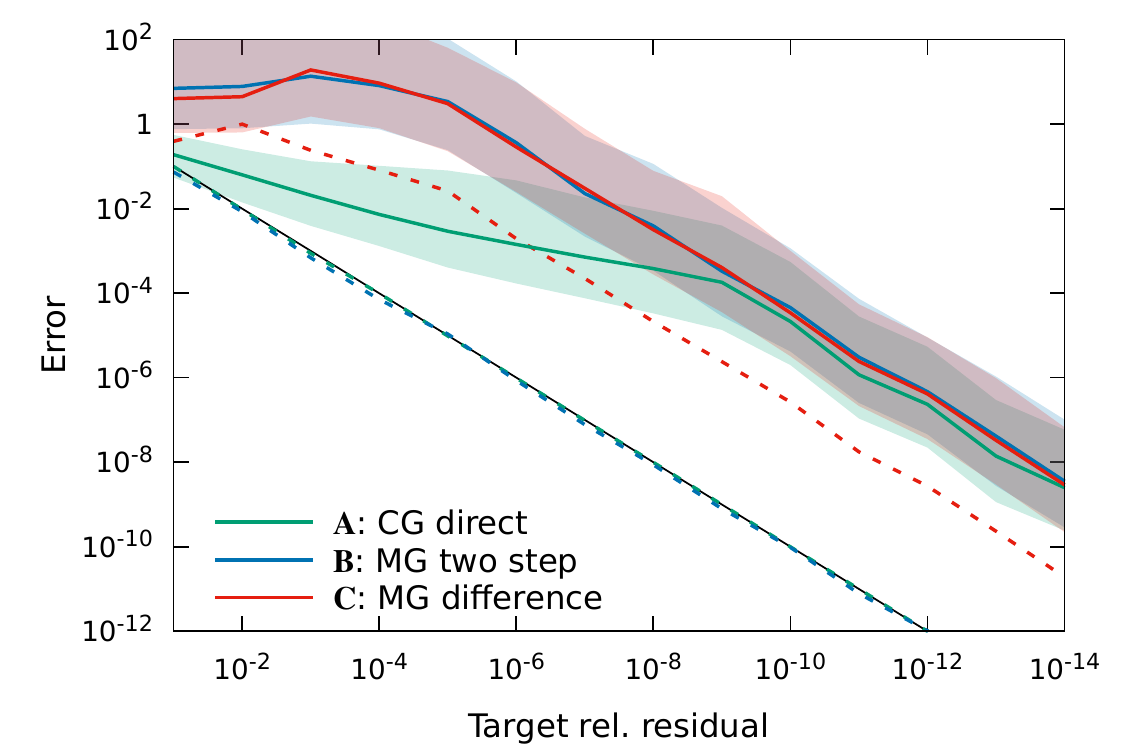}
	\caption{The dependence of the error and the residual on the stopping criterium for
	the three different ways of solving Eq.~\eqref{eq:sqlineq}. Method \textbf{A} uses the conjugated gradient solver (green),
	method \textbf{B} uses  the consecutively ordered single inversions of Eq.~\eqref{eq:linmul} (blue) and
	method \textbf{C} uses the single inversions involving differences (red) given by Eq.~\eqref{eq:lindif}.
	The residuals are shown as dashed lines while the error is illustrated by the solid line with a shaded 
	band determined by the largest and smallest local deviations as defined in Appendix~\ref{sec:error}.}
	\label{fig:error_square}
\end{figure}

In the case of TM Wilson fermions at maximal twist
this relation is fulfilled since $\lambda_{min} \sim 0$.
This is the case in our example, shown in Fig.~\ref{fig:error_square}.
CG solver and the two aforementioned multigrid approaches are analyzed using a stopping criteria of $\norm{r}/\norm{b}<10^{-9}$
for our physical test ensemble. For this ensemble, $\lambda_{min}^2+\mu^2 \simeq 0.00072^2$, thus
yielding a difference around six orders of magnitude between residual and error, as shown in Fig.~\ref{fig:error_square}.  

Despite the fact that as long as $\mu \gg \lambda_{min}$ all approaches yield a similar error,
the approach \textbf{C} has advantages.
The first advantage is that the software optimization
is straightforward because both shifts can be solved together, as
outlined in Ref.~\cite{Heybrock:2015kpy,Richtmann:2016kcq}. The second advantage is that for the shifted linear equations
it gives a better way to start the iteration with an optimal initial
guess. Therefore, we consider this approach in what follows.

\subsection{DD-$\alpha$AMG for the non-degenerated twisted mass operator}\label{sec:DDalphaAMG_for_1p1TM}

The idea behind  adapting the DD-$\alpha$AMG approach 
to the ND twisted mass operator $D_{\ND}(\bar\mu,\bar\epsilon)$ is based on  preserving the $(\Gamma_{5}\otimes \tau_1)$-symmetry on the coarse grid.
We define the ND coarse grid operator by 
\begin{equation}
D^{\phantom\dagger}_{\ND,c}(\bar\mu,\bar\epsilon) = P^\dagger_{\ND}D^{\phantom\dagger}_{\ND}(\bar\mu,\bar\epsilon)P^{\phantom\dagger}_{\ND}.
\end{equation}
with $P_{\ND}$ being a suitable prolongation operator.
If $P_{\ND}$ is $(\Gamma_5\otimes \tau_1)$-compatible, i.e.
\begin{equation}\label{eq:g5-t1-compatible}
(\Gamma_5\otimes \tau_1)P_{\ND} = P_{\ND}(\Gamma_{5,c}\otimes \tau_1),
\end{equation}
it follows that the $(\Gamma_5\otimes \tau_1)$-hermiticity of $D^{\phantom\dagger}_{\ND}$
in Eq.~(\ref{eq:g5tau1-hermiticity}) is also preserved on the coarse grid
and the coarse grid operator fulfills
\begin{equation}\label{eq:g5-t1-hermiticity}
(\Gamma_{5,c}\otimes \tau_1)\,D_{\ND,c}^{\phantom\dagger} 
= P^\dagger_{\ND}(\Gamma_5\otimes \tau_1)\,D^{\phantom\dagger}_{\ND}P^{\phantom\dagger}_{\ND}
= P^\dagger_{\ND}D^{\dagger}_{\ND}(\Gamma_5\otimes \tau_1)P^{\phantom\dagger}_{\ND}
= D_{\ND,c}^{\dagger}(\Gamma_{5,c}\otimes \tau_1)\,.
\end{equation}
The property in Eq.~(\ref{eq:g5-t1-compatible}) is satisfied by a prolongation operator $P_{\ND}$, which is  $\Gamma_5$-compatible and diagonal 
in flavor space. We choose identical components in flavor space defining $P_{\ND} = P \otimes I_2$.
Thus, we obtain
\begin{equation}\label{eq:non-degenerate_coarse}
D_{\ND,c}(\bar\mu,\bar\epsilon)= (D_{\W,c}\otimes I_2)+i\bar{\mu}\,(\Gamma_{5,c}\otimes\tau_3)-\bar{\epsilon}\,(I_c\otimes\tau_1)=
\begin{bmatrix}
D_{\TM,c}(\bar\mu)        & -\bar\epsilon\,I_c \\
-\bar\epsilon\,I_c & D_{\TM,c}(-\bar\mu),
\end{bmatrix} 
\end{equation}
which follows from the property  $P^\dagger P = I_c$.
We note that the flavor and spin components of the coarse operator preserve
a similar sparse structure and properties of the fine grid operator 
$D_{\ND}(\bar\mu,\bar\epsilon)$ in Eq.~(\ref{eq:non-degenerate}).

It follows from Eq.~(\ref{eq:error_propagation}), that the prolongation operator $P$
has to project onto a subspace, which captures the IR-modes. While $P_{\ND}$ 
is degenerate in flavor space, the low modes of the ND twisted mass operator
are defined in the full space. Thus our solution to Eq.~(\ref{eq:g5-t1-compatible})  i.e.~$P_{\ND}=P \otimes I_2$  could spoil the efficiency of the coarse grid
correction since the same prolongation operator $P$ has to act on both flavor spaces.
A possible solution is to use the  prolongation operator $P$   employed
for the TM Wilson operator~\cite{Alexandrou:2016izb}. 
This has the advantage that if the multigrid solver is used during the HMC 
the same setup, built up for one flavor of the TM Wilson operator, can be reused in any step of the HMC
for both light degenerate and heavy non-degenerate sector.
Saving additional setup expenses makes the usage of the multigrid in HMC more effective.
We will motivate this choice in section~\ref{sec:motivation_MG1+1},
and discuss some numerical results in section~\ref{sec:speed-up_MG1+1}.

\subsubsection{Motivations for the multigrid construction}\label{sec:motivation_MG1+1}
\begin{figure}
	\includegraphics[width=\textwidth]{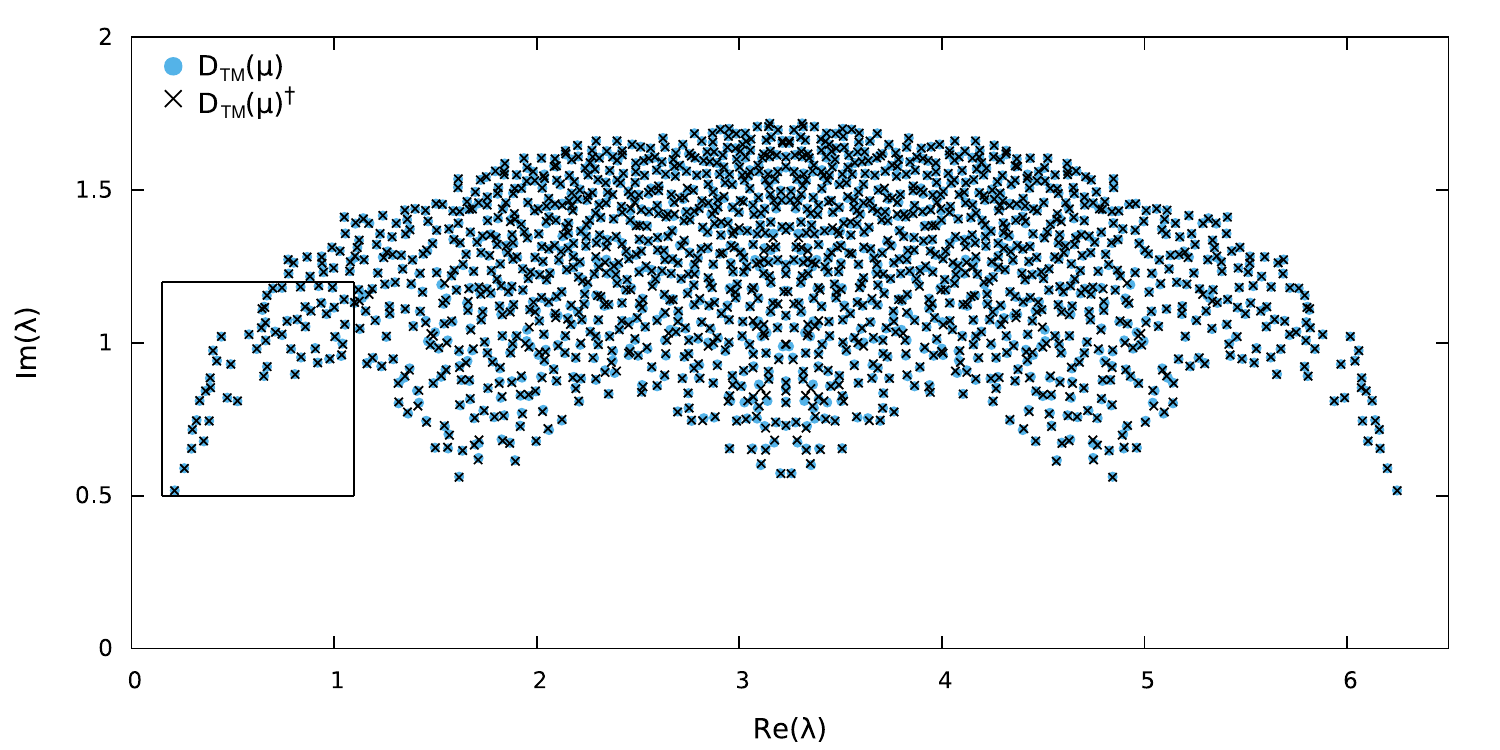}
	\includegraphics[width=\textwidth]{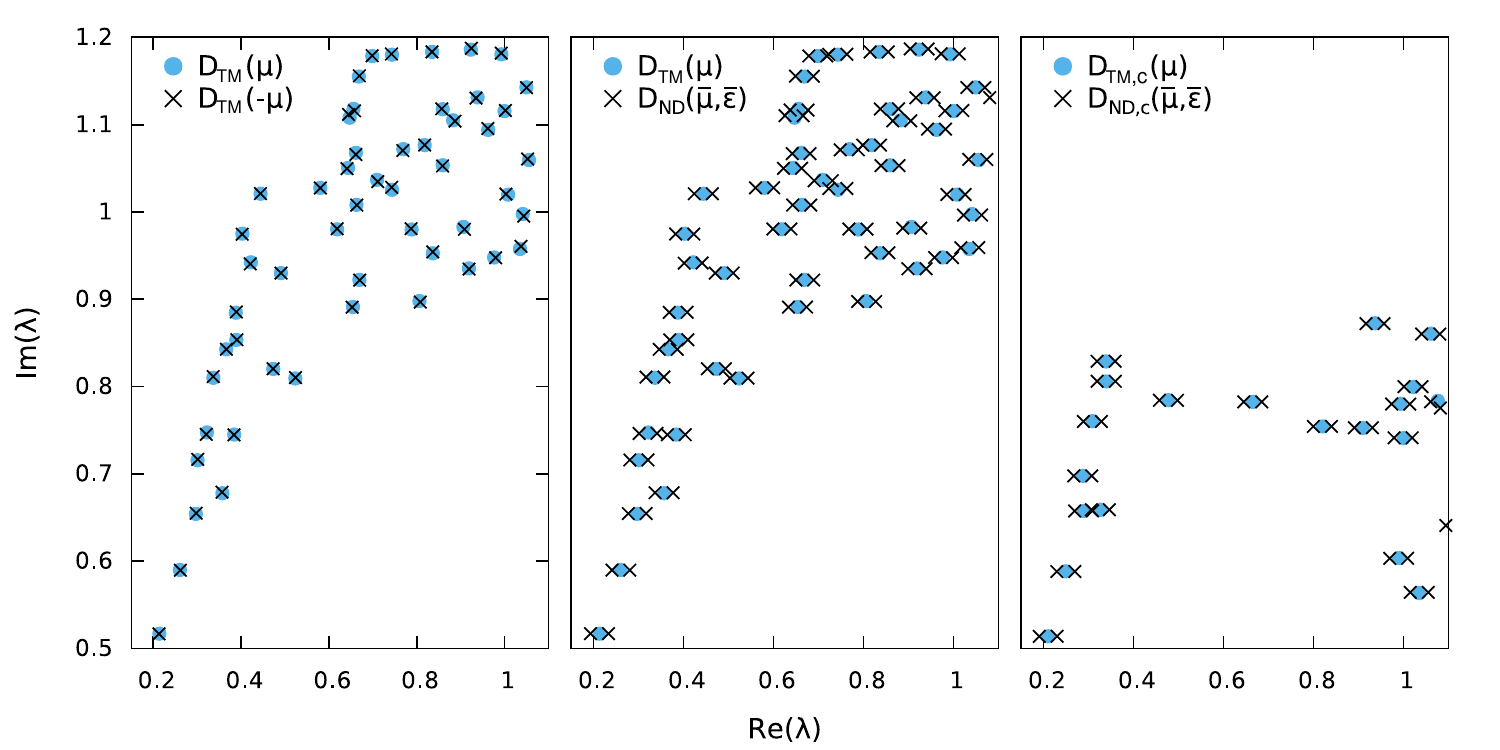}
	\caption{\label{fig:non-degenerate_spectrum}Complete spectrum of a gauge configuration for a lattice of size $4^4$.
	The twisted mass parameters are $\mu,\bar\mu,\bar\epsilon=0.01$.
	In the top panel, we depict the spectrum of the degenerate $D_\TM(\mu)$ operator folded with 
	respect to the imaginary axis. In the bottom left panel we show part of the spectrum 
	close to the origin for $D_\TM(\mu)$ for different sign of $\pm\mu$. In the central lower panel
	we compare the degenerate and non-degenerate TM operator. In the  
	right lower panel, we show the spectrum for the coarse version of both operators.
	}
\end{figure}

The choice of the subspace $P_{\ND} = P \otimes I_2$ 
for Eq.~(\ref{eq:non-degenerate_coarse}) is motivated as follows:
\begin{itemize}
	\item Let us consider first the case with $\bar\epsilon=0$, for which the subspace needs to be
	effective for the TM Wilson operator $D_{\TM}(\pm\mu)$ where $\mu$ can take both sign.
        This was analyzed numerically in Ref.~\cite{Alexandrou:2016izb,Bacchio:2016bwn} where it was found that
        the effectiveness of the subspace generated for the TM Wilson operator 
        $D_{\TM}(\mu)$ is affected only slightly if another parameter set is used
	with $D_{\TM}(\pm\rho)$ where $\rho\ge\mu$. 
	
	This can be explained by the connection between 
	right-handed eigenvectors $v_R$ of the TM Wilson operator $D_{\TM}(\mu)$
	and the left-handed eigenvectors $v_L=v_R^\dagger \Gamma_5$ of $D_{\TM}(-\mu)$, which reads as
	\begin{equation}
	D_{\TM}(\mu) v_R = \lambda  v_R \quad \iff \quad v_R^\dagger \Gamma_5 D_{\TM}(- \mu)  = v_R^\dagger \Gamma_5 \lambda^\dagger .
	\end{equation}
	Thanks to the $\Gamma_5$ compatibility of the aggregation, this is also true for the coarse operator.
	Thus, the eigenspaces of $D_{\TM}(\pm\mu)$ are connected and a prolongation operator $P$ constructed for
	$D_{\TM}(\mu)$ captures the low modes of $D_{\TM}(\mu)$ when acting on the right while acting from the left givens the low modes of $D_{\TM}(-\mu)$.
	
	\item Considering the case $D_{\ND}(0,\bar\epsilon)$ where $\bar\mu=0$ and $\bar\epsilon \neq 0$,
	it follows that the eigenvalues have a linear dependency in $\bar\epsilon$ and the eigenvectors are degenerate in flavor space.
	Indeed the relation $D_{\ND}(0,\bar\epsilon) w_{\pm} = (\lambda \pm \bar\epsilon) w_{\pm}$ with $w_\pm = (v, \pm v)$ holds in flavor space 
	with the eigenvector $v$ of $D_\W$ satisfying $D_\W v = \lambda v$.
	Thus, for $\bar\mu = 0$ the eigenspace is invariant under changing $\bar\epsilon$
	and it motivates the choice via Eq.~\eqref{eq:g5-t1-compatible} for the coarse grid projector.
	
	\item It follows that the choice for the projector of Eq.~\eqref{eq:g5-t1-compatible}
        is well-motivated for the special cases $\bar\mu \neq 0 \, \wedge\, \bar\epsilon = 0$ and  $\bar\mu=0 \, \wedge\, \bar\epsilon \neq 0$.
        However, what we have in reality is $\bar\mu\sim 0.1 \, \wedge\, \bar\epsilon \sim 0.1$. In order to verify
        if the properties mentioned in the previous two items are a good approximation, we study numerically the eigenvalue spectrum when
        changing the parameters $\bar\mu$ and $\bar\epsilon$ on a small lattice of a size of $4^4$.
        The eigenvalue spectrum with positive imaginary part of the ND twisted mass operator
        is depicted in Fig.~\ref{fig:non-degenerate_spectrum}. Note that due to the $(\Gamma_5\otimes\tau_1)$-hermiticity
        the real-axis is a symmetry axis, thus eigenvalues come in complex conjugated pairs or are real. For the one flavor operator $D_\TM(\mu)$ this symmetry
        is broken for $\mu \neq 0$. This is shown in the left  lower panel,  where we focus on the spectrum of the low modes.
        By taking a closer look to the spectrum of the ND twisted mass operator, the real-axis symmetry is restored.
        Moreover, the parameter $\bar\epsilon$ for the non-degenerate operator $D_{\ND}(\bar\mu,\bar\epsilon)$ acts on
        the eigenvalues via a linear shift $\pm\bar\epsilon$ similar to the ideal case given by $D_{\ND}(0,\bar\epsilon)$.
        Thus, the projector given by Eq.~\eqref{eq:g5-t1-compatible} should project on the small eigenvalues of $D_{\ND}(\bar\mu,\bar\epsilon)$. 
        For our case study, this is indeed the case
        as depicted in the right panel of Fig.~\ref{fig:non-degenerate_spectrum}. The spectrum of the coarse grid operator for the  TM Wilson and ND twisted mass operator display similar  features 
        showing that preserving the operator structure in the coarse grid allows to preserve properties of the fine operator.
\end{itemize}

\subsubsection{Numerical results}\label{sec:speed-up_MG1+1}

\begin{figure}
	\centering
	\includegraphics[width=0.49\textwidth]{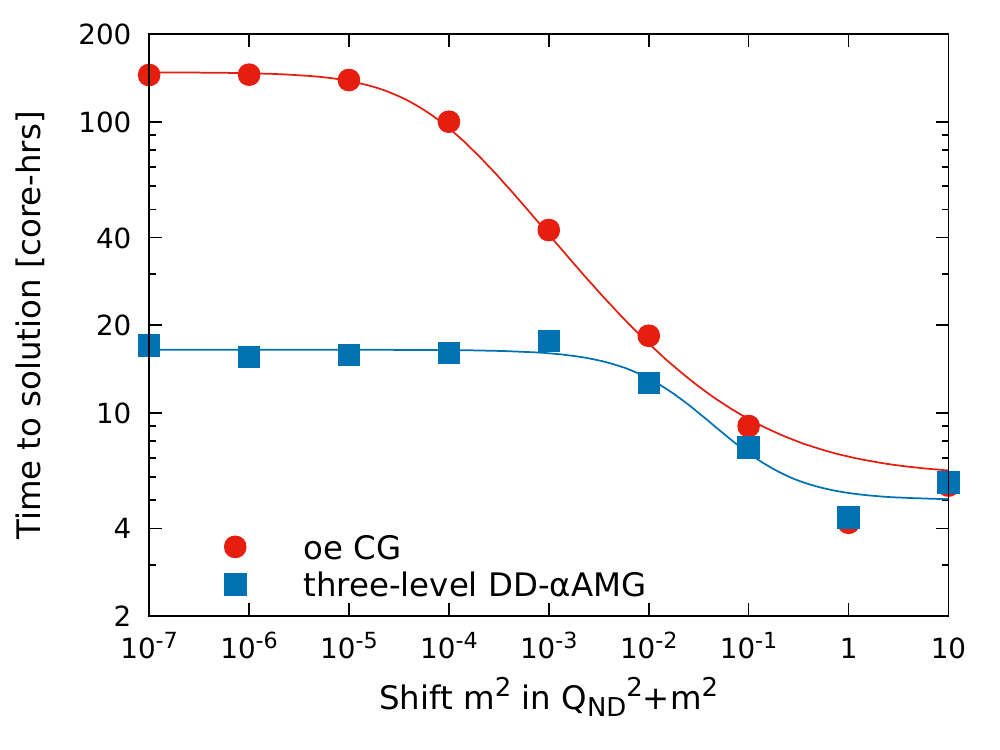}
	\includegraphics[width=0.49\textwidth]{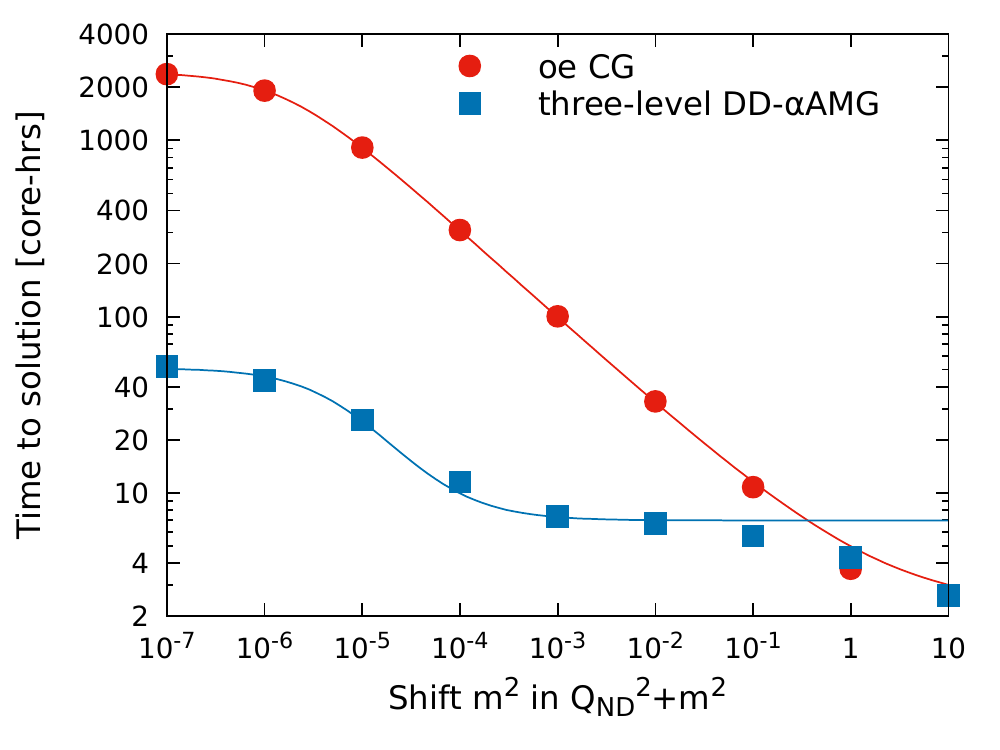}
	\caption{Comparison between time to solution for computing the inverse of the squared ND
	twisted mass operator at different shift $m^2$ using the odd-even (oe) CG solver and the DD-$\alpha$AMG approach.
	We used the physical test ensemble employing physical strange and charm quark masses (left panel) and
	physical up and down quark masses (right panel).}
	\label{fig:crit_slow_down_1+1}
\end{figure}

In this section we test the effectiveness of our choice $P_\ND = P \otimes I_2$ using
the physical test ensemble.
We compare the performance of the multigrid solver using method \textbf{C}, see Eq.~\eqref{eq:lindif},
to the one of the conjugate gradient
solver in the case of the squared ND twisted mass operator at physical
heavy quark parameters and additional at physical light quark parameters.
Solving the equation
\begin{equation}
  x = (Q_{\ND}^2(\bar\mu,\bar\epsilon)+m^2)^{-1} b  = \frac{i}{2m} (Q_{\ND}(\bar\mu,\bar\epsilon)+ im)^{-1} b - (Q_{\ND}(\bar\mu,\bar\epsilon)- im)^{-1} b)
  \label{eq:sqND}
\end{equation}
for several shifts in $m^2$ we obtain the results shown in Fig.~\ref{fig:crit_slow_down_1+1}.
The time to solution for the CG solver can be fitted by~\cite{saad2003iterative}
\begin{equation}\label{eq:CGtime2}
 t_{CG}(m^2) = a + \frac{b}{\textrm{ln} \left( 1-2/(\sqrt{\kappa(m^2)}+1) \right)}
\end{equation}
with the condition number $\kappa = (\lambda_{max}+m^2)/(\lambda_{min}+m^2)$.
In the case of the heavy quark parameters it follows
$\lambda_{max}=4.7$ and $\lambda_{min}=0.000065$ and 
at minimal $\chi^2$ the parameters are given by $a=5.88(123)$ and $b=-1.06(3)$.
The fit approach used for the time to solution of the multigrid solver
is motivated by the convergence of the general minimal residual solver (GMRES)~\cite{LieTic04b}. We employ the
functional form given by
\begin{equation}\label{eq:MGtime2}
 t_{MG}(m^2)=A+\frac{B}{\textrm{ln}(1-(C+m^2)/(4.7+m^2))}
\end{equation}
where we use, instead of the condition number, a modified function dependence given by $(C+m^2)/(4.7+m^2)$.
Minimizing $\chi^2$ for the heavy quark parameters yields $A=5.0(7)$, $B=-0.059(26)$ and $C=0.024(10)$.
Note that this is only an effective fit approach, which describes the data well but
could fail for different cases. At the physical strange and charm quark masses 
-- i.e.~$m^2\rightarrow 0$ see left panel of Fig.~\ref{fig:crit_slow_down_1+1} --
we found an order of magnitude speed-up of the multigrid approach compared to the CG solver.
Moreover, the ND multigrid solver is even more effective than the multigrid solver for the TM Wilson operator
at physical strange quark mass. This can be seen by comparing the relative speed-up 
for strange quark mass of $m_q\sim 95 \; \textrm{MeV}$, shown in Fig.~\ref{fig:crit_slow_down} for the TM Wilson operator.
The relative speed-up for two application of the ND-multigrid solver is comparable with the speed-up of a single application
in case of the TM Wilson operator. 
This is also confirmed
at physical non-degenerated light quark masses, as shown in the right panel
of Fig.~\ref{fig:crit_slow_down_1+1}. Here we found a speed-up of around two orders of
magnitude similar to the case of the TM Wilson operator.
This shows that the choice of the coarse grid projector,
building up using the TM Wilson operator, yields
a very effective multigrid approach for the ND twisted mass operator.

Due to the large parameter set of multigrid approaches, optimization 
for a specific lattice can become a major task. In Ref.~\cite{Alexandrou:2016izb}
we outlined our strategy and gave a set of parameters.
We use this set of optimized parameters  with a few adjustments in case of heavy quark masses.
Namely, the shift $d$ of the TM parameter in the coarse grid 
is set to unity  and the number of smoothing iterations
is reduced from 4 to 2.

\begin{figure}
	\centering
	\includegraphics[width=0.7\textwidth]{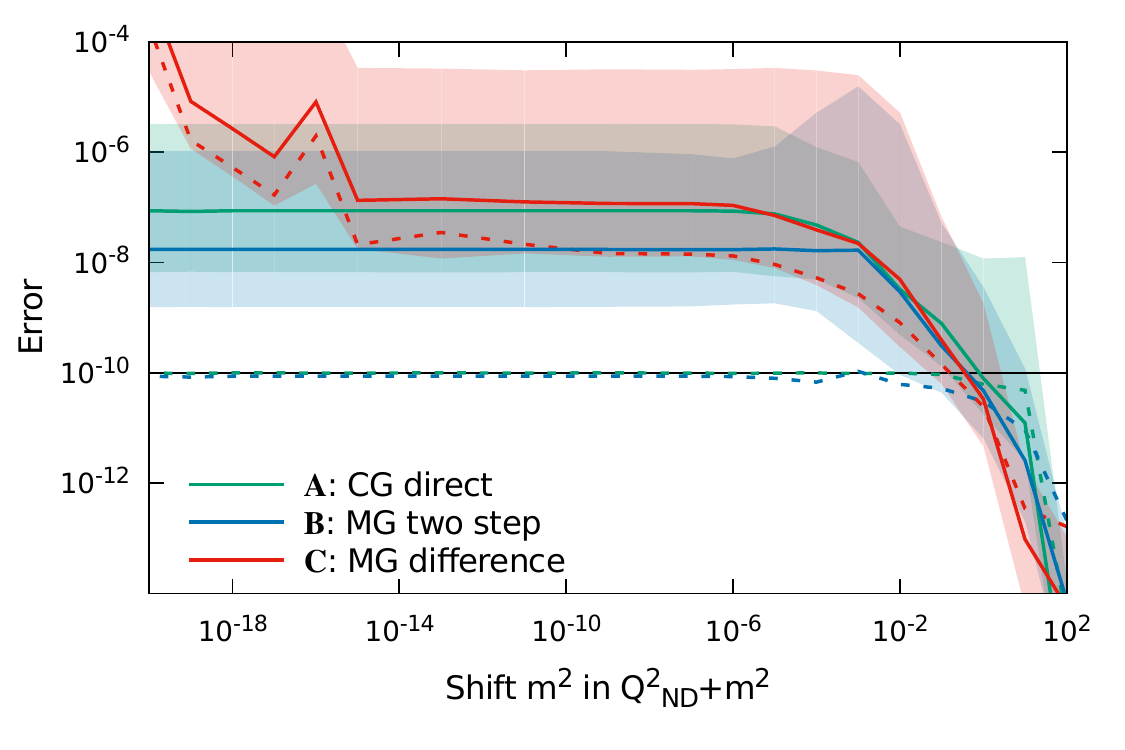}
	\caption{The  dependence of the error and the residual on the shift $m^2$ for the three different approaches for solving
	the linear equation involving the ND twisted mass operator. Method A is given by the direct solution using
	the conjugate gradient solver (green), method B is given by the consecutively ordered single solves (blue)
	and method C is given by the difference of the two single solves (red). The residual are shown as dashed lines
	while the error is illustrated by the straight line attached to a shaded band given by the largest and smallest deviations
	as defined in Appendix \ref{sec:error}.}
	\label{fig:error_square_nd_shift}
\end{figure}

As discussed in  section~\ref{sec:sqsol},  Eq.~\eqref{eq:sqND} with the squared operator can be
solved in different ways. We analyze the connection of the error of the squared system with the residual of
single systems for different shifts $m$ at physical heavy ND parameters $\bar \mu$ and $\bar \epsilon$. The results are
shown in Fig.~\ref{fig:error_square_nd_shift}. 
We fix the stopping criteria for each inversions to $\norm{r}/\norm{b}<10^{-10}$.
As discussed in section~\ref{sec:sqsol} and in  Appendix~\ref{sec:error}, they depend
on the smallest eigenvalue of the operator $Q^2_{\ND}+m^2$, which is given by
$\lambda_{\ND,min}\sim 0.0008$.
As shown in Fig.~\ref{fig:error_square_nd_shift} this bounds the
error for the approach \textbf{A} and \textbf{B} by around 3 orders of magnitudes
if $m<\lambda_{\ND,min}$. For case \textbf{C} this is not true, however, numerically we find
that the real error is similar to case \textbf{A} and \textbf{B}  when $m^2 > 10^{-14}$.
This shows that method \textbf{C} can be used for our application since in the rational approximation
of the matrix square root all shifts are larger than $m^2>10^{-10}$.

\section{Multigrid in rational approximation}\label{sec:MGinit_guess}

In this section, we introduce a novel strategy for providing initial guesses to
shifted linear systems as in Eq.~(\ref{eq:shifted_linear}) in order
to optimize the usage of multigrid approaches in multi mass-shifted systems.
We first give a brief introduction to the rational approximation
of the square root of a hermitian matrix before we  discuss
how optimal initial guesses can be generated.

\subsection{Rational approximation of the square root}

Rational functions can be used to approximate analytical functions
like the square root of a matrix, which is an essential ingredient 
for lattice QCD. For example, in case of the Rational Hybrid Monte Carlo (RHMC) \cite{Clark:2006fx}, which is commonly
used in the case of Wilson-type fermions for non-degenerate quark masses, staggered fermions and  for calculating the sign-function needed
for the Neuberger overlap operator.
Here we are interested in solving
\begin{equation}
 x = \sqrt{Q^{-2}} b~.
 \label{eq:sqND1}
\end{equation}
In general a continuous function, like the square root, can be approximated by a
rational function of generic order $[m,n]$
\begin{equation}\label{eq:rational_n_m}
R(y)=A\, \frac{\prod_{i=1}^{m} (y+n_i)}{\prod_{i=1}^{n} (y+d_i)}~.
\end{equation} 
The maximal deviation in a fixed interval of this rational approximation is then 
bounded from below, as stated in the de-Vall\'ee-Poussin's theorem~\cite{achieser2013theory}.
A rational approximation is optimal when the maximal deviation is equal to the bound.
In case of the function $1/\sqrt{y}$ an optimal rational approximation is
given by Zolotarev's solution~\cite{zolotarev1877application}.
The rational function of order $[n,n]$
\begin{equation}\label{eq:rational}
R_{n,\epsilon}\left(y\right) = a_{n,\epsilon} \prod_{j=1}^{n}\frac{y+c_{n,\epsilon,(2j-1)}}{y+c_{n,\epsilon,2j}}
\end{equation}
optimally approximates $1/\sqrt{y}$
in the interval $\epsilon<y<1$ with a maximal deviation
\begin{equation}\label{eq:deviation}
\delta_{n,\epsilon} = \max_{\epsilon<y<1}\abs{1-\sqrt{y}R_{n,\epsilon}\left(y\right)}~.
\end{equation}
The parameters in Eqs.~(\ref{eq:rational},\ref{eq:deviation}) can be computed analytically \cite{Luscher:2010ae} and they are given by
\begin{align}
c_{n,\epsilon,k} &= \cs^2\left(k \cdot v_{n,\epsilon},\sqrt{1-\epsilon}\right) &\text{with }\ v_{n,\epsilon} = \frac{K\left(\sqrt{1-\epsilon}\right)}{2n+1}\label{eq:zolotarev_shifts}\\
a_{n,\epsilon} &= \frac{2}{1+\sqrt{1-d_{n,\epsilon}^2}}\prod_{j=1}^{n}\frac{s_{n,\epsilon,(2j-1)}}{s_{n,\epsilon,2j}}&\text{with }\ s_{n,\epsilon,k} = \sn^2\left(k \cdot v_{n,\epsilon},\sqrt{1-\epsilon}\right)\\
\delta_{n,\epsilon} &= \frac{d_{n,\epsilon}^2}{1+\sqrt{1-d_{n,\epsilon}^2}}&\text{with }\ d_{n,\epsilon} = (1-\epsilon)^{\frac{2n+1}{2}}\prod_{j=1}^{n}s_{n,\epsilon,(2j-1)}^2
\end{align}
where $\sn(u,k)$ and $\cs(u,k)=\cn(u,k)/\sn(u,k)$ are Jacobi elliptic functions and $K(k)$ is the complete elliptic integral.

The solution to the Eq.~\eqref{eq:sqND1} with fixed $\epsilon$ and order $n$
can be rewritten as a system with a sum over operators with multiple mass shifts
given by \begin{equation}\label{eq:ratapp}
 x \simeq R_{n,\epsilon}(Q^2) b = a_{n,\epsilon} \left(1 + \sum_{i=1}^{n} q_i  \cdot(Q^2+m_i^2)^{-1}\right) b
\end{equation}
where
\begin{equation}
m_i^2=c_{n,\epsilon,2i} \qquad  \textrm{and} \qquad q_i= (c_{n,\epsilon,(2i-1)}-c_{n,\epsilon,2i}) \prod_{j=1, j\neq i}^{n}\frac{c_{n,\epsilon,(2j-1)}-c_{n,\epsilon,2i}}{c_{n,\epsilon,2j}-c_{n,\epsilon,2i}}~. 
\end{equation}

\subsection{Initial guesses for shifted linear systems}\label{sec:init_guess}

Iterative solvers are initiated via a starting vector. In most of the cases,
this vector is chosen to be zero, since it is the most safe starting point
if the inverse is unknown. However, if parts of the inverse is known
the iteration count can be minimized by starting with an initial
vector close to the solution, e.g.~this is done in the case of exact
deflation where the smallest eigenvalues are used to
preconditioning the system. Here, we  discuss
two approaches, proposing an initial guess based on previous solutions
and via the MMS-CG solver.

\subsubsection{Initial guesses via Lagrange interpolation}

The idea for the case of multi-mass shifted systems is
to use previously computed solutions, 
e.g.~$x_{1}, \; \ldots , x_{n}$, to generate an initial guess for the next inversion of the solution $x_{n+1}$.
This can be done by a polynomial interpolation of the previous solutions where the previous shifts enter as node points.
A polynomial of degree $(n-1)$, which interpolates $n$ solutions is then given by
\begin{equation}
 p(m) = \sum_{i=1}^{n} l_{i,n}(m) x_i \quad\text{with }\ l_{i,n}(m) = \prod_{\substack{j=1\\j\neq i}}^{n} \frac{m-m_j}{m_{i}-m_{j}},
\end{equation}
where $l_{i,n}$ are the Lagrange polynomials.
The initial guess for the $(n+1)$th system follows with $\tilde{x}_{n+1} = p(m_{n+1})$.

Let us compute the $n$ solutions using the stopping criteria $\norm{r_i}/\norm{b} < \rho$,
then an upper bound for the initial guess for the $(n+1)$th iteration is given by 
\begin{equation} \label{eq:init_guess_res_lim_1}
\norm{\tilde{r}_{n+1}} = \norm{b-(Q+m_{n+1} I)\tilde{x}_{n+1}} \le \rho\,\gamma_{n}\norm{b} +\norm{\sum_{i=1}^{n} c_{i,n} (m_{n+1}-m_{i})\,x_i}, 
\end{equation}
where $\gamma_n$ is the sum over the absolute values of the coefficients $c_{i,n} = l_{i,n}(m_{n+1})$. More details are given in Appendix \ref{sec:proof}.

The second term of Eq.~\eqref{eq:init_guess_res_lim_1} can be approximated
assuming $x_i = (Q+m_i)^{-1} b$ such that it can be cast into 
\begin{equation}\label{eq:init_guess_res_lim_2}
\frac{\norm{\tilde{r}_{n+1}}}{\norm{b}} \lesssim \gamma_n\rho + \prod_{i=1}^{n} \frac{\abs{m_{n+1} - m_i}}{\lambda_{\min}+m_i}
\end{equation}
if $m_i>0$ and $Q$ a positive-(in)definite matrix with $\lambda_{\min}\geq0$.
The initial guess $\tilde{x}_{n+1}$ is bounded by two terms.
The first term depends on the nodes $m_i$ while the second 
term depends additionally on the smallest eigenvalue of the matrix.
It follows that $\norm{\tilde{r}_{n+1}}/\norm{b}> \rho $ since
$\gamma_n > 1$. However, $\gamma_n$ is known {\it at priori}
such that the interpolation strategy can be adapted if $\gamma_n$ becomes too larger.
In our case the second term dominates the right hand side of Eq.~\eqref{eq:init_guess_res_lim_2}
due to the dependence on $\lambda_{min}$. Moreover the second term is strictly smaller than one for the case
$m_{n+1} < m_i$. This gives the optimal ordering to solve the 
multi-shifted problem via initial guesses with
\begin{equation}
 m_1 > m_2 > \ldots > m_n > m_{n+1} > \ldots > m_N > 0,
\end{equation}
whereby it follows
\begin{equation}\label{eq:init_guess_res_lim_3}
0\leq \prod_{i=1}^{n} \frac{\abs{m_{n+1} - m_i}}{\lambda_{\min}+m_i}<1.
\end{equation}
Finally, we  remark that if the upper bound of Eq.~\eqref{eq:init_guess_res_lim_2}
is smaller than 1, starting from $\tilde{x}_{n+1}$ will be always more efficient than
from a zero vector.

\subsubsection{Initial guesses via MMS-CG}

Another possibility to guess a starting vector for the
last $(N-n)$ shifts is given by using the MMS-CG solver.
The general idea of the MMS-CG solver is to exploit the fact 
that the eigenspace of the shifted systems are
identical. Thus the Krylov space generated for
one of the shifts can be simultaneously used to iterate
the other shifts. 

If one generates the Krylov space
for the most ill-conditioned system, here the $N$th system,
all other iteration vectors will converge to smaller residuals than the residual of the target system.
However, if the system is too ill-conditioned, like it is in our
case, the iteration count of the MMS-CG solver increases
drastically and a hybrid-approach is potentially  more efficient.

Our proposal is to solve the first $n$ systems via the MMS-CG solver
by generating the Krylov space for the $n$th system. This will generate
the first $n$ solutions $x_1,\; x_2,\ldots, x_n$. 
Furthermore, the MMS-CG solver can also predict guesses for the next $m$ systems
by iterating those together with the first $n$ systems. Although the iteration vectors
of these systems will not reach the target precision, at the iteration count
where the $n$th system is converged, the iteration vectors will contain
the full information of the generated Krylov space for the $n$th system.
Based on this fact,  for the $(n+1)$th system, the MMS-CG solver generates an initial
vector, which is in general closer to the target residual than using a Lagrange interpolation
based on the $n$ solutions.

\subsection{Initial guesses for the rational approximation of the square root}\label{sec:init_guess_RHMC}

In the following we will analyze the behavior of the initial guesses for the rational approximation of the
square root in Eq.~\eqref{eq:ratapp} for the ND twisted mass operator using the physical test ensemble
at physical strange and charm quark masses.
For that we choose a rational approximation
consisting of 10 terms using the interval $(\epsilon_\ND \;;\; 1]$
with $\epsilon_\ND=0.000065/4.7$.
Using Lagrange interpolation we obtain an upper bound
for the initial guesses through Eq.~\eqref{eq:init_guess_res_lim_1},
bounded by two terms, which depend on $\gamma_{n}(m)$ 
and the previous solutions $x_i$, respectively.
The coefficient $\gamma_{n}(m)$ depends on the nodes of the
interpolation given by the shifts $m_i^2$. Using an ordering
$m_1>m_2> \ldots >m_N$ for the shifts we found for both cases
that $\gamma_n(m_{n+1})$ is not larger than $2$. This bounds 
the first term of Eq.~\eqref{eq:init_guess_res_lim_1} by $2 \rho$ where 
$\rho$ is the precision of the stopping criterion. Thus, 
 the first term is suppressed compared to the second term of Eq.~\eqref{eq:init_guess_res_lim_1}
such that we can neglect it in the following.

\begin{figure}
	\centering
	\includegraphics[width=0.7\textwidth]{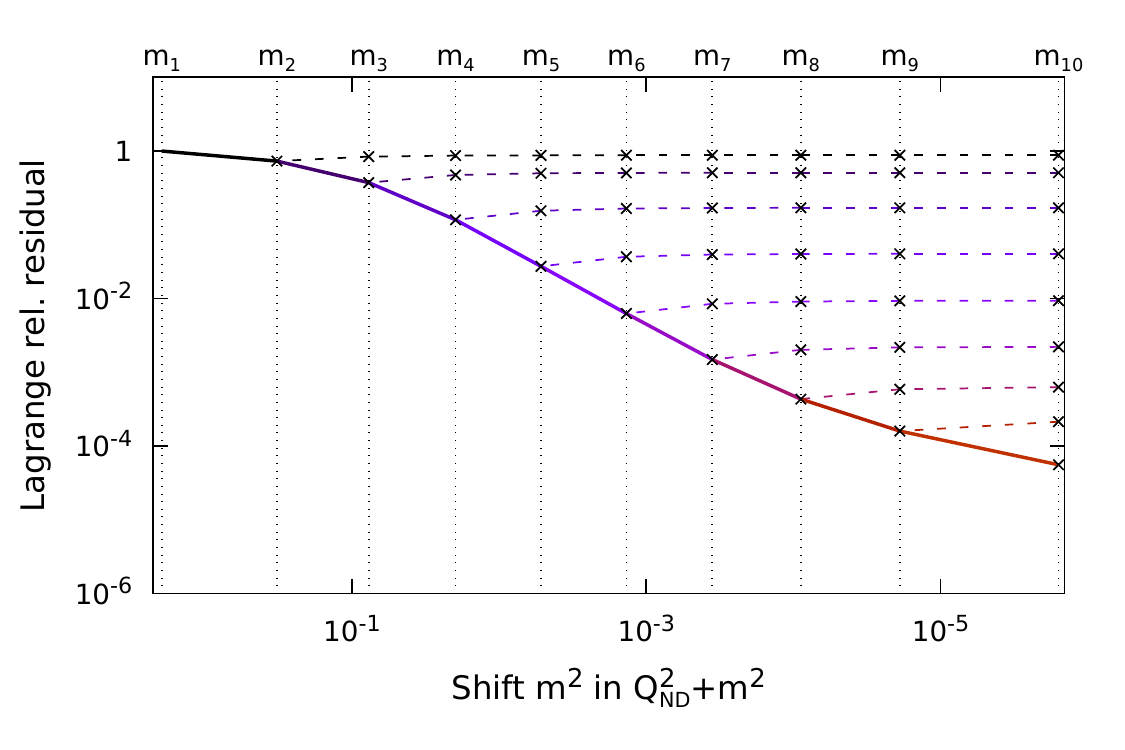}
	\caption{ The  relative residuals of the Lagrange interpolation based on the $n$ solutions 
	versus the shifts $m^2$ used in the approximation of the square root of the ND twisted mass operator.
	The solid lines illustrate the norm of the relative residual of the initial guess 
	for decreasing $m$ taking all available nodes into account. 
	The nodes used in the interpolation, i.e. the shifts $m^2$, are denoted by vertical dotted lines. We depict the dependence of  each polynomial $p^{(i)}$
	for $m^2>m_i^2$ by the dashed lines.}
	\label{fig:inigues_ndlag}
\end{figure}

The second term of Eq.~\eqref{eq:init_guess_res_lim_1} can be approximated via 
\begin{equation}
 \norm{\sum_{i=1}^{n} c_{i,n} (m_{n+1}-m_{i})\,x_i} \simeq  \prod_{i=1}^{n} \frac{\abs{m_{n+1} - m_i}}{\lambda_{\min}+m_i}
\end{equation}
with $\lambda_{\min} \sim 0.0008$ for the squared ND twisted mass operator.
Using an ordering $m_1>m_2> \ldots >m_N$ this yields
an upper bound of $0.0003$ for the last, the $N$th, initial guess,
which is close to the residual shown in Fig.~\ref{fig:inigues_ndlag}.
However, for the first shifts the real residual is around one magnitudes lower
than this bound. We find that the data can be described effectively via
\begin{equation}
 \norm{\sum_{i=1}^{n} c_{i,n} (m_{n+1}-m_{i})\,x_i} \lesssim  \prod_{i=1}^{n} \frac{\abs{m_{n+1} - m_i}}{A_i m_i},
\end{equation}
where $A_i$ is for all $i$ smaller then $1.9$. Using $B=[\textrm{max}(A_i)]^{-1}<1$ it follows
\begin{equation}
 \norm{\sum_{i=1}^{n} c_{i,n} (m_{n+1}-m_{i})\,x_i}  \simeq  B^n \prod_{i=1}^{n} \left(1-\frac{m_{n+1}}{m_i}\right)
\end{equation}
with $m_{n+1}<m_i$ and $B\cong 0.6$.
Thus, the initial guesses using the Lagrange interpolation become more efficient with increasing $n$,
as shown in figure Fig.~\ref{fig:inigues_ndlag}.

\begin{figure}
	\centering
	\includegraphics[width=0.7\textwidth]{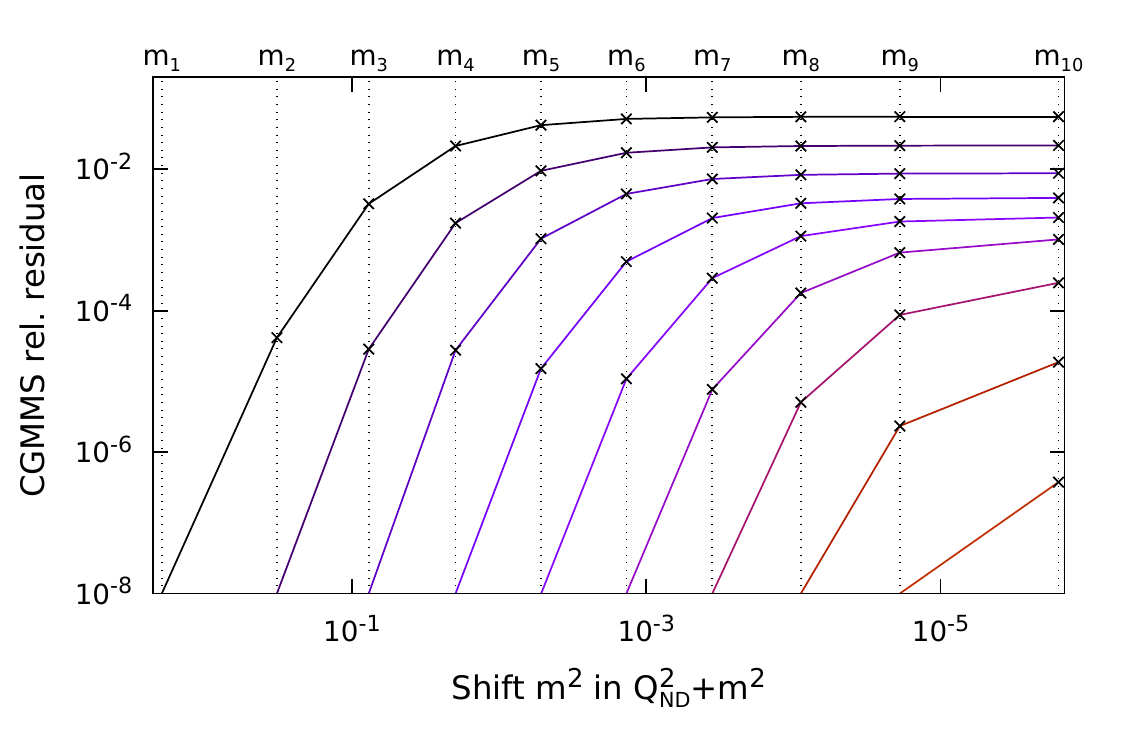}
	\caption{The  relative residual of the iteration vectors
	of the MMS-CG versus the mass shifts when varying the target Krylov
	space for the ND twisted mass case. The solver is stopped when the residual
	of the target system reaches the stopping criterion given here by $\norm{r}/\norm{b}<10^{-8}$.
	The iteration vectors of systems with smaller shift than the target one have a 
	larger residual.  This is shown for all $N-1$ cases depicted by the solid line
	changing the color from black, for the largest target shift
	to orange for the $N-1$ shift. Note that we do not show the case where the target system is given by
	the last, $N$th system, because here all other systems have converged.
	The vertical dotted lines illustrate the shifts $m_i^2$.}
	\label{fig:init_guess_4_1p1}
\end{figure}

As pointed out in the previous section, the MMS-CG solver,
used for solving the first $n$ systems,
can be used to predict an initial guess by including
the $(N-n)$ systems in the MMS-CG iteration. This effectively
interpolates the initial vector $x_j$ in the Krylov space
of the $n$th system. We depict the residual for the predicted
initial guesses by the MMS-CG solver in Fig.~\ref{fig:init_guess_4_1p1}
if the $n$th system is converged. As shown, the relative residual of the 
$(n+1)$th system depends only slightly on $n$ and is given by $\sim 10^{-5}$.
Thus for the first step after using the MMS-CG solver the residual is smaller then the residual of 
the initial guess generated by the Lagrange interpolator. However,
this changes for $m>1$. While for small $n$ the guess
using the MMS-CG solver for the $(n+2)$th system is better, 
for $n \gtrsim 5$ the guesses of the Lagrange interpolation 
yield similar results and becomes better for $m>2$.

Based on these results, the optimal approach for our example is a combination of
all three elements, namely the MMS-CG solver and the Multigrid solver
with initial guesses using the MMS-CG solver and the Lagrange interpolation.
Thus, solving the first $n$th systems 
using  MMS-CG solver involves an additional iteration for the $(n+1)$th and $(n+2)$th systems,
which can be used as an initial guess. The $(N-n)$ remaining systems are 
then solved via a multigrid approach one by one where for the last $(N-n-2)$ systems the Lagrange 
interpolation is used to start the iteration. In the following section,
we will discuss the optimal $n$.

\subsection{Results}\label{sec:results}

Based on the observations in the previous section, we propose to use a hybrid approach
to solve a system of linear equations with $N$ shifts. Namely, 
use the MMS-CG solver for the largest $n$ shifts and the multigrid approach for the remaining $(N-n)$ systems solving each one
via the difference methods discussed in section~\ref{sec:sqsol}. The $(N-n)$ systems
can be started by initial guesses, proposed for the $(n+1)$th and $(n+2)$th systems using 
the MMS-CG solver and for the rest using Lagrange interpolation and employing the previous solutions.

The optimal $n$ depends on the ratio of the performance of the MMS-CG
solver and the multigrid approach, which includes environmental parameters,
software implementation and computer hardware. Here, we consider the Haswell-nodes
partition of SuperMUC and use an MPI parallelization employing 1024 task
on 37 nodes. The software used is the tmLQCD
package~\cite{Jansen:2009xp,Finkenrath:tmLQCD}, which is linked to the DDalphaAMG library~\cite{Rottmann:DDalphaAMG} and is publicly available.

The  question we would like to answer is what is the optimal
$n$ of the  hybrid approach introduce in this work, i.e.~how many $n$ shifts should be solved
with the MMS-CG solver in order to solve the total system with $N$ shifts in
an efficient way employing for the last $N-n$ shifts a multigrid approach.
This we  discuss by employing the physical test ensemble using the ND twisted
mass operator at the strange and charm quark masses with 10 mass-shifts. We employ 
case \textbf{C}, the difference method, for the linear equation with the squared operator when the multigrid solver is used.

\begin{figure}
	\centering
	\includegraphics[width=0.7\textwidth]{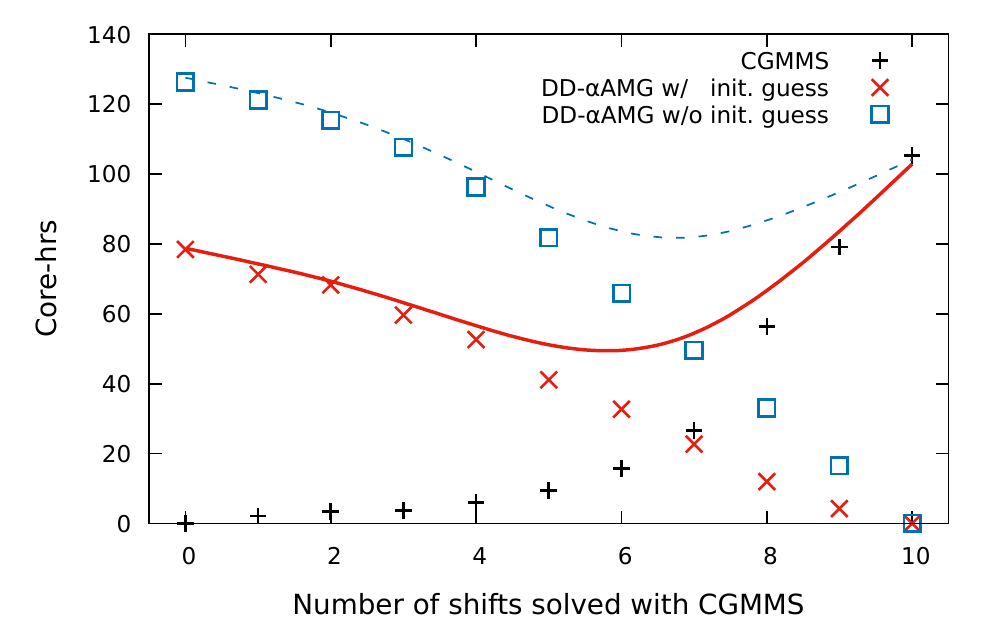}
	\caption{The  CPU hours necessary to solve the multi-mass shifted equation
	involving the ND twisted mass operator at the strange and charm quark masses versus the number of shift $n$ used in the MMS-CG solver shown as black crosses. The left $10-n$ shifts
	are solved via the multigrid approach with initial guesses as red crosses and without initial guesses as blue squares.
	The total CPU hours needed to solve the system is shown by the red solid line for the hybrid method
	using initial guesses and for the hybrid method without initial guesses with the red solid line.}
	\label{fig:CGvsMG_strange}
\end{figure}

The cost for the solution of the multi-mass shift linear equation via the hybrid method is given by
\begin{equation}\label{eq:costwo}
 c_{HY}(n) = t_{CG}(m_n^2) + \sum_{i=n+1}^N t_{MG}(m_i^2)~.
\end{equation}
where the time to solution of the MMS-CG solver is approximated with the time of CG solver $t_{CG}(m)$ at the smallest shift $m_n$.
For the case without initial guesses, the cost $c_{HY}(n)$ can be minimized using the fits Eqs.~\eqref{eq:CGtime2} 
and \eqref{eq:MGtime2}. This yields an optimal $n_{opt} \cong 7$ as shown in Fig.~\ref{fig:CGvsMG_strange}.
However, the total speed-up of the hybrid method without initial guesses only improves
slightly the time to solution compared to the application of the MMS-CG solver.
This changes by using initial guesses. Here, the cost for the multigrid part
is significantly reduced as shown in Fig.~\ref{fig:CGvsMG_strange} using initial guesses compared to the case without.
We find that the initial guesses reduce the total time to solution by about factor of two  
while the optimal $n_{opt} \sim 6$ is shifted slightly.

The effect of this improvement can be also seen in the ensemble
generation~\cite{Alexandrou:2018,Bacchio:2017pcp}, used in the HMC.
When the force terms of the rational approximation are split to integrate shifts on different
time scales, the  time to solution of the heavy quark sector is reduced by a factor of approximately two.
Note that the setup comes without additional cost due to the usage of the setup generated for the light quarks
as discussed in section \ref{sec:DDalphaAMG_for_1p1TM}.

\begin{figure}
	\centering
	\includegraphics[width=0.7\textwidth]{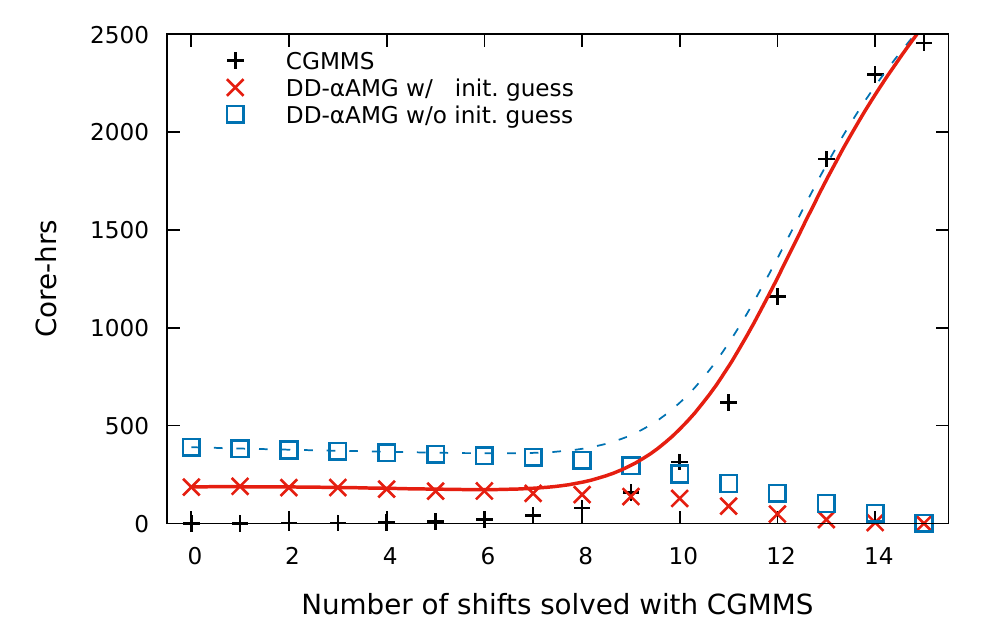}
	\caption{The  CPU hours necessary to solve the multi-mass shifted equation
	involving the ND twisted mass operator for the physical value of the up and down quark mass. The dependence
	on the number of shift $n$ used in the MMS-CG solver is shown as black crosses. The left $15-n$ shifts
	are solved via the multigrid approach and results  with initial guesses are shown with the red crosses while without initial guesses with the blue squares.
	The total CPU hours needed to solve the system is given by the red solid line for the hybrid method
	using initial guesses and with the  dotted blue line for the hybrid method without initial guesses.}
	\label{fig:CGvsMG_up}
\end{figure}

The hybrid method becomes even more effective for smaller quark masses. To illustrate this,
the $n$-dependence is examined using the ND twisted mass operator at physical
light quarks using the physical test ensemble.
For $\bar \mu_\ell=0.00072$ and $\bar \epsilon_\ell=0.000348$  the dependence
of the hybrid method using a rational approximation with $N=15$ terms is shown in Fig.~\ref{fig:CGvsMG_up}.
We find that the hybrid method with initial guesses gives a total speed up at optimal $n_{opt} \cong 7$ 
by approximately a factor of 15. As in the case of the strange and charm quark masses,  the initial
guesses, result in a speedup of about a factor of two.

\section{Conclusions}\label{sec:conclusions}

In this paper we examine  multi-mass shifted system
involving the ND twisted mass operator.
We discuss in detail how the time to solution can be minimized
using a hybrid method based on the MMS-CG solver in combination with
a multigrid approach by using initial guesses.
Using the heavy flavor doublet, and tuning to the physical values of the strange
and charm quark masses, we find that employing a hybrid approach
a speed up by a factor two can be achieved using for
the six largest shifts the MMS-CG solver
and for the smallest shifts the multigrid approach
in combination with initial guesses.
We show that for the light quark doublet this
hybrid approach yields even a larger improvement speeding up the time to solution  
by a factor of about 15.

The  approach presented  also speeds up the force calculation
of the current $N_f=2+1+1$ simulations of the ETM collaboration.
Furthermore, this approach can be applied for all lattice fermion
discretization for which a multigrid approach exists and a rational
approximation is the choice to generated an ensemble of gauge configuration,
e.g.~like it is the case for the staggered discretization scheme.

One essential tool to speed up the hybrid method for the value of the strange
quark mass is the usage of initial guesses. In this paper
we show that a combination based on prediction via the MMS-CG solver
and an approach based on Lagrange interpolation of the previous
solutions yields the best time to solution.

We also discuss in detail the DD-$\alpha$AMG approach, showing
how to preserve symmetries of the fine grid operator on the coarse
grid. Preserving the $(\Gamma_5\otimes \tau_1)$-hermiticity of the
ND twisted mass operator by employing a coarse grid projection
based on the TM Wilson operator yields a stable multigrid method,
yielding an overall speed-up of one magnitude compared to the CG solver
at physical values of the strange and charm quark masses. This shows
that the DD-$\alpha$AMG approach is very effective as long as
important symmetries are preserve on the coarse grid.

Furthermore, different ways for solving 
a linear equation with a shifted squared Dirac operator using iterative
solvers are presented. While the CG solver can be used to solve this system directly the 
multigrid approach can be only applied to a system,
which involves one Dirac operator. This results into two possible
ways of solving the system, one using two consecutively solves and one using differences
of single solutions. What has been shown in this work is that the latter can be used
in the case of the ND twisted mass operator at physical strange and charm quark masses 
and yields some advantages if initial guesses are used.

\subsection*{Acknowledgments}
This project has received funding from the  Horizon 2020 research and innovation programme of the European Commission
under the Marie Sklodowska-Curie grant agreement No 642069. S.B.~is supported by this programme.
We would like to thank Andreas Frommer, Roberto Frezzotti, Carsten Urbach and Bartosz Kostrzewa for fruitful discussions.
The authors gratefully acknowledge the Gauss Centre for Supercomputing e.V. (www.gauss-centre.eu) for funding the project \emph{pr74yo}
by providing computing time on the GCS Supercomputer 
SuperMUC at Leibniz Supercomputing Centre (www.lrz.de).

\bibliography{Multigrid_in_RHMC}

\appendix

\section{Error and residual of equivalent solutions}\label{sec:error}

In this appendix we compare the numerical error and residual of solutions 
which are equivalent in exact arithmetic but not numerically.
For clarity sake, our notation is the following: a linear system 
\begin{equation}\label{eq:linA}
A x = b,
\end{equation} 
with $A$ being an invertible matrix, $b$ a known right hand side (rhs) and $x$ the numerical solution,
has residual and error, respectively,
\begin{equation}\label{eq:res_error}
r = b - A x \quad \text{and} \quad e =  A^{-1} b-x.
\end{equation}
The residual is commonly used as stopping criterium for the solvers,
since it does not require the knowledge of the \textit{exact solution} $A^{-1} b$.
However the error vector $e$ is the statistical deviation which one introduces using the solution $x$.
The error is connected to the residual by the relation
\begin{equation}\label{eq:error_norm}
e =  A^{-1} r \quad \text{which states} \quad \norm{e} = \norm{A^{-1} r} \le \norm{A^{-1}} \norm{r} = \frac{\norm{r}}{\sqrt{\lambda_{\min}(A^\dagger A)}} ,
\end{equation}
where the latter equality holds in the Euclidean norm and $\lambda_{\min}(A^\dagger A)$ is the smallest eigenvalue of $A^\dagger A$.
Iterative solvers are usually stopped when $\norm{r}<\rho\norm{b}$ with $\rho$ being a
fixed tolerance. This fix the relative norm of the residual to be below a given threshold.

Hereafter we study and show numerical results for equivalent solutions
in several cases. We use the TM Wilson operator on the physical test ensemble. In the numerical results we also show the error with
maximal and minimal deviation of the local components.
In the following examples we start from a random vector taken as the solution $x$.
Then we apply the involved operator $A$ to obtain the right hand side $b=A x$. 
Then the error is given by $e = x' -x$ with $x'$ the iterated solution. 
The minimal and maximal local deviation are given min/max of the norm of $e$
restricted to the lattice site.
The maximal and minimal deviation are potentially interesting for future simulations where very large lattices will 
be used~\cite{Luscher:2017cjh}. Indeed since $\norm{b}$ grows with $V^{1/2}$,
the stopping criterium $\norm{r}\le\rho\norm{b}$ does not guarantee the
residual or error to be uniformly small and at larger $V$ larger local
deviation of the vectors are allowed.

\subsection{Squared twisted mass operator}

We start by taking a closer look to the linear system $(Q^2 + \mu^2 I)x=b$
which involves a squared hermitian operator $Q^2$. As already discussed in Section \ref{sec:sqsol} we consider here the
following three methods:
\begin{itemize}
	\item[\textbf{A:}] the linear system is directly solved by
	\begin{equation}\label{eq:direct}
	x_A+ e_A = (Q^2+\mu^2I)^{-1}b \,.
	\end{equation}
	The solver stopping criterium is based on the relative residual of the solution, thus
	\begin{equation}
	\norm{r_{A}} = \norm{b-(Q^2 + \mu^2 I)x_{A}}  \le \rho\norm{b}
	\end{equation}
	and norm of the error satisfies
	\begin{equation}
	\norm{e_{A}} = \norm{(Q^2 + \mu^2 I)^{-1}r_{A}} \le \frac{\norm{r_{A}}}{\lambda^2_{\min} + \mu^2} \le \frac{\rho\norm{b}}{\lambda^2_{\min} + \mu^2}~.
	\end{equation}
	
	\item [\textbf{B:}] the system is solved in two consecutive steps, by computing 
	\begin{equation}\label{eq:two-step}
	x_{\pm}+ e_{\pm}=(Q\pm i\mu I)^{-1} b  \quad \text{and then } \quad x_B + e'_B=(Q\mp i\mu I)^{-1}x_{\pm}
	\end{equation}
	using either $x_{+}$ or $x_{-}$. The solution
	\begin{equation}\label{eq:two-step-error}
	x_B+ e_B= (Q^2+\mu^2I)^{-1}b  \quad \text{has error } \quad e_B = e'_B + (Q\mp i\mu I)^{-1}e_{\pm}.
	\end{equation}
	The solver stopping criteria are in the two steps respectively
	\begin{align}
	\norm{r_{\pm}} &= \norm{b-(Q\pm i\mu I)x_{\pm}}  \le \rho\norm{b}\label{eq:res-pm}\\
	\norm{r'_{B}} &= \norm{x_{\pm}-(Q\mp i\mu I)x_{B}} \le \rho\norm{x_{\pm}} \leq \frac{\rho\norm{b}}{\sqrt{\lambda^2_{\min} + \mu^2}}.
	\end{align}
	The residual of the solution is then
	\begin{align}
	r_B &= b- (Q^2+\mu^2I)x_B = b- (Q\pm i\mu I)(x_{\pm} + r'_B) = r_{\pm} + (Q\pm i\mu I) r'_B\nonumber\\
	&\Longrightarrow  \norm{r_B} < \norm{r_{\pm}} + \norm{Q\pm i\mu I} \norm{r'_B} \le \left(1+\sqrt{ \frac{\lambda^2_{\max} + \mu^2}{\lambda^2_{\min} + \mu^2}}\right)\rho\norm{b}.\label{eq:two-step-res}
	\end{align}
	For obtaining the smallest upper limit of the error in Eq.~(\ref{eq:two-step-error}), we consider
	\begin{align}
	e_{\pm} = (Q\pm i\mu I)^{-1}r_{\pm,B} \quad&\Longrightarrow\quad \norm{e_\pm} \le \frac{\rho \norm{b}}{\sqrt{\lambda^2_{\min} + \mu^2}}\label{eq:error-pm}\\
	e'_B = (Q\mp i\mu I)^{-1}r'_B \quad&\Longrightarrow\quad \norm{e'_B} \le \frac{\rho \norm{x_{\pm}}}{\sqrt{\lambda^2_{\min} + \mu^2}} \le \frac{\rho\norm{b}}{\lambda^2_{\min} + \mu^2}
	\end{align}
	from which we obtain
	\begin{equation}
	\norm{e_B} = \norm{e'_B + (Q\mp i\mu I)^{-1}e_{\pm}} \le \norm{e'_B} + \frac{\norm{e_{\pm}}}{\sqrt{\lambda^2_{\min} + \mu^2}} \le \frac{2\rho\norm{b}}{\lambda^2_{\min} + \mu^2}.\label{eq:two-step-err}
	\end{equation}
	
	\item [\textbf{C:}] the solution is given by a difference of two solutions of a linear combination of $x_{\pm}+ e_{\pm}=(Q\pm i\mu)^{-1}b $, 
	\begin{equation}\label{eq:linear-combination}
	x_C = b - (Q^2 + \mu^2) x_C = \frac{i}{2\mu}(x_+-x_-) = (Q^2+\mu^2I)^{-1}b - \frac{i}{2\mu}(e_+-e_-) \qquad\text{for }\mu \neq 0.
	\end{equation}
	The solver stopping criteria are in Eq.~(\ref{eq:res-pm}) and the errors $e_{\pm}$ in Eq.~(\ref{eq:error-pm}).
	The residual of the solution is then
	\begin{align}
	r_C &= \frac{i}{2\mu}\left((Q+i\mu I)r_+-(Q-i\mu I)r_-\right)\nonumber\\
	\Longrightarrow  \norm{r_C} &\le \frac{1}{2\abs{\mu}}\left(\norm{Q+i\mu I}\norm{r_-} + \norm{Q-i\mu I}\norm{r_+}\right) \le  \frac{\rho\sqrt{\lambda^2_{\max} + \mu^2}\norm{b}}{\abs{\mu}} \label{eq:linear-combination-res}
	\end{align}
	while the error of the solution is
	\begin{align}
	e_C  = \frac{i}{2\mu}(e_+-e_-)\quad\Longrightarrow\quad  \norm{e_C} \le \frac{1}{2\abs{\mu}} (\norm{e_+}+\norm{e_-}) \le \frac{\rho\norm{b}}{\abs{\mu}\sqrt{\lambda^2_{\min} + \mu^2}}.\label{eq:linear-combination-err}
	\end{align}
\end{itemize}

Most of the numerical results are discuss in Section \ref{sec:sqsol} and depicted in Fig.~\ref{fig:error_square} for the $N_f=2$ TM Wilson operator 
and in Fig.~\ref{fig:error_square_nd_shift} for the ND twisted mass operator.
Here we want to make some additional remarks which are the following ones:
\begin{itemize}
      \item The error bounds of method \textbf{A}, \textbf{B} and \textbf{C} are compatible as long as the shift $\mu$ is larger than $\lambda_{\min}$.
	  If the shift becomes smaller the error bound of method \textbf{C} increases inverse proportional. In our numerical example
	  we found for all methods a comparable error, but 
	  only for the case $\mu<10^{-14}$ with $\lambda_{\min} \sim 0.0008$ for the ND twisted mass case
	  we found a deviation using method \textbf{C}. Here, methods \textbf{A} or \textbf{B} yield better results.
      \item The difference between the real error and the error bounds can give some information on the modes the solver
	    tackles effectively. Namely if the real error coincidences with the error bound,
	    the error is dominated by the mode of the smallest eigenvalue. In contrast, if the error is much smaller than
	    the error bound, the error is dominated by much larger modes and thus the solver would treat the small eigenmodes
	    very effectively. In our numerical test we found that in all cases the smallest eigenvalues are dominating the
	    error. This is also the case if a multigrid solver is used. An explanation is that the coarse grid
	    correction is calculated with a very large stopping criteria, which just tackle the low modes
	    until the fine grid precision is reached.
 	
	\item The error norm has a slope parallel to the target relative residual when the residual vector is dominated by a specific eigenvector.
 	Indeed $\norm{e} = \norm{(Q^2 + \mu^2 I)^{-1}r} = \alpha \norm{r}$ only if $r$ is an eigenvector.
 	In general the residual of standard Krylov solvers is dominated by the eigenvector with smallest eigenvalue which in this case is $\mu^2$.
 	This is indeed the separation we observe in Fig.~\ref{fig:error_square} between error and relative target residual.
 	Interestingly also multigrid methods produce solutions which have a residual dominated by the smallest eigenvalue.
 	It is surprising because multigrid methods threat the low mode subspace separately and the convergence is expected to be similar for all the modes.
 	For instance an exactly deflated solver would have the residual dominated by the first non-deflated eigenvalue.
 	Moreover the multigrid residual is dominated by the smallest eigenvalue since target relative residual $\rho>10^{-5}$ returning a solution with 100\% error.
 	This also explain why for inversions at high target relative residual, as in the molecular dynamics trajectory, 
 	we need with the DD-$\alpha$AMG solver a more accurate 
 	solution than for the CG solver to satisfying the reversibility check at the same precision, as reported in Ref.~\cite{Bacchio:2017pcp}.
\end{itemize}

\subsection{Single flavor twisted mass operator}
\begin{figure}
	\centering
	\includegraphics[width=0.7\textwidth]{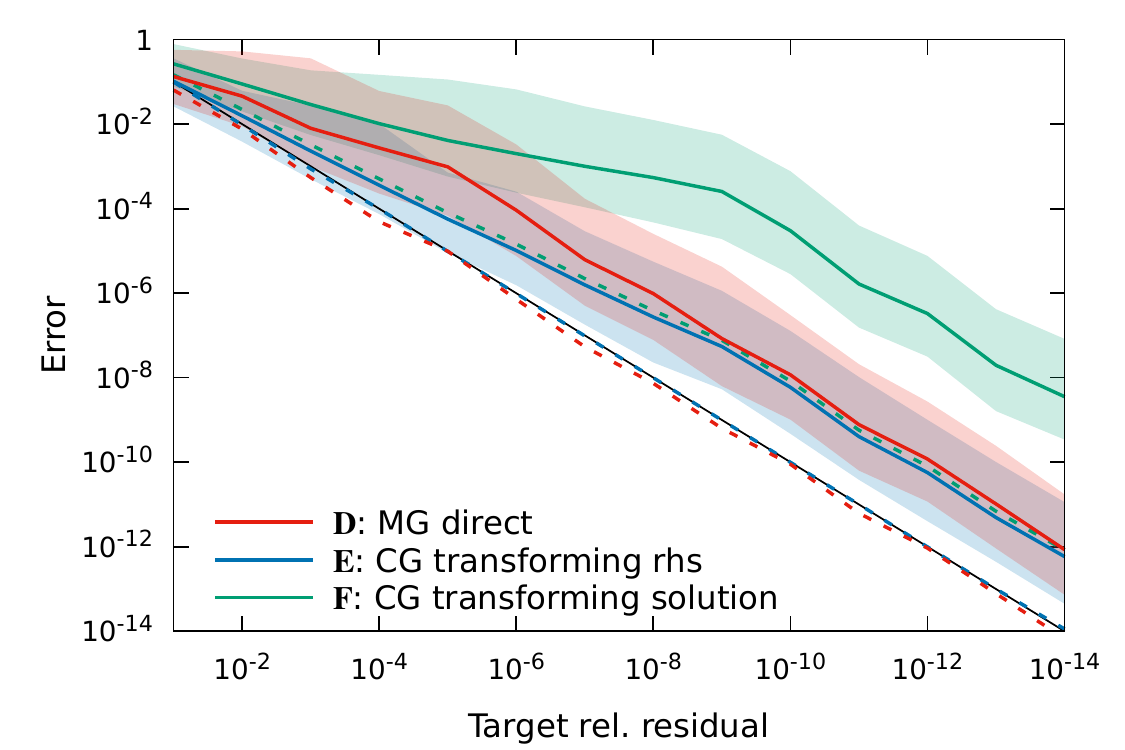}
	\caption{The  dependence of the error and the residual on the stopping criteria for the three different approaches for solving
	the linear equation involving one single TM Wilson operator. Method \textbf{D} is given by the direct solution using
	the multigrid solver (red), method \textbf{E} is given by the CGNR method (blue)
	and method \textbf{F} is given by the CGNE method (green). The residual are shown as dotted lines
	while the error is illustrated by the straight line attached to a shaded band given by the largest and smallest deviations.}
	\label{fig:error_plus}
\end{figure}

In general retrieving the solution of 
the linear system in Eq.~(\ref{eq:linA}) can be done by different methods, for instance by using a linear iterative solver like 
the conjugate gradient (CG) solver.
In case of CG, which requires $A$ to be hermitian instead of Eq.~\eqref{eq:linA}
a modified linear equation involving a hermitian operator has to be solved.
Two methods are available, known as CGNR and CGNE given by solving the equations
\begin{equation}
\quad(A^\dagger A) x = A^\dagger b\quad \textrm{or} \quad(A A^\dagger) y =  b \rightarrow x = A^\dagger y
\end{equation}
for obtaining $x$ respectively.
Although both solutions are equivalent in exact arithmetic, in practice the numerical
error and residual, as defined in Eq.~\eqref{eq:res_error}, are in general different.
In order to approach smaller quark masses and increasing the dimension of the lattice require to study different ways
in order to control the resulting error.

Here we consider the following three methods for solving the linear equation $(Q\pm i\mu I)x_\pm=b$:
\begin{itemize}
	\item[\textbf{D:}] The linear system is solved directly by obtaining
	\begin{equation}
	x_{\pm,D}+ e_{\pm,D}=(Q\pm i\mu I)^{-1}b .
	\end{equation}
	The solver stopping criterium is based on the relative residual of the solution, thus
	\begin{equation}
	\norm{r_{\pm,D}} = \norm{b-(Q\pm i\mu I)x_{\pm,D}} \le \rho\norm{b}.
	\end{equation}
	Following Eq.~(\ref{eq:error_norm}), the norm of the error satisfies
	\begin{equation}
	\norm{e_{\pm,D}} = \norm{(Q\pm i\mu I)^{-1}r_{\pm,D}} \le \frac{\norm{r_{\pm,D}}}{\sqrt{\lambda^2_{\min} + \mu^2}} \le \frac{\rho\norm{b}}{\sqrt{\lambda^2_{\min} + \mu^2}}
	\end{equation}
	where $\lambda^2_{\min}$ is the smallest eigenvalue of $Q^2$.

	\item[\textbf{E:}] A normal equation is used by applying a transformation on the rhs (equivalent to CGNR)
	\begin{equation}
	x_{\pm,E}+ e_{\pm,E}=(Q^2 + \mu^2 I)^{-1}(Q\mp i\mu I)b .
	\end{equation}
	Since the rhs is $(Q\mp i\mu I)b$, the solver stopping criterium is
	\begin{equation}
	\norm{r'_{\pm,E}} = \norm{(Q\mp i\mu I)b - (Q^2+ \mu^2 I)x_{\pm,E}} \le \rho \norm{(Q\mp i\mu I)b} \le \rho\norm{b}\sqrt{\lambda^2_{\max} + \mu^2}.
	\end{equation}
	The norm of the residual is then
	\begin{equation}
	\norm{r_{\pm,E}} = \norm{(Q\mp i\mu I)^{-1}r'_{\pm,E}} \le \frac{\rho\norm{(Q\mp i\mu I)b}}{\sqrt{\lambda^2_{\min} + \mu^2}} \le \rho \norm{b} \sqrt{\frac{\lambda^2_{\max} + \mu^2}{\lambda^2_{\min} + \mu^2}}
	\end{equation}
	and the norm of the error satisfies
	\begin{equation}\label{eq:error-single-B}
	\norm{e_{\pm,E}} \le \frac{\norm{r_{\pm,E}}}{\sqrt{\lambda^2_{\min} + \mu^2}} \le \frac{\rho \norm{b} \sqrt{\lambda^2_{\max} + \mu^2}}{\lambda^2_{\min} + \mu^2}.
	\end{equation}
	
	\item[\textbf{F:}] A normal equation is used by applying a transformation on the solution (equivalent to CGNE)
	\begin{equation}
	y_{F} + e_{F}=(Q^2 + \mu^2 I)^{-1}b \quad\longrightarrow\quad x_{\pm,F} + e_{\pm, F}=(Q\mp i\mu I) y_{F} 
	\end{equation}
	where $e_{\pm, F} = (Q\mp i\mu I) e_{F}$.
	The solver stopping criterium
	\begin{equation}
	\norm{r_F} = \norm{b-(Q^2+ \mu^2 I)y_{F}} = \norm{b-(Q \pm \mu I)x_{\pm,F}} = \norm{r_{\pm,F}} \le \rho\norm{b}
	\end{equation}
	is equivalent to computing the residual of the solution $r_{\pm,F}$. The norm of the error then satisfies
	\begin{equation}
	\norm{e_{\pm,F}} \le \frac{\norm{r_{\pm,F}}}{\sqrt{\lambda^2_{\min} + \mu^2}} \le \frac{\rho\norm{b}}{\sqrt{\lambda^2_{\min} + \mu^2}}.
	\end{equation}
\end{itemize}

From this analysis, we conclude that method \textbf{D} and \textbf{F} generates
solutions which have compatible residuals and errors. On the other hand, 
method \textbf{E}  has upper limits increased by the condition 
number $\kappa=\sqrt{(\lambda^2_{\max} + \mu^2)/(\lambda^2_{\min} + \mu^2)}$ compared to \textbf{D} or \textbf{F}. 
The numerical results depicted in Fig.~\ref{fig:error_plus} for the $N_f=2$ twisted mass operator at 
the physical light quark mass ($\lambda^2_{\min} = 0$ and $\mu = 0.00072$) confirm these conclusions. 
From the numerical results we also notice that the error
of the method \textbf{E} is close to the upper limits obtained in Eq.~(\ref{eq:error-single-B}). 
This shows that the residual of CG solver is dominated by the lowest eigenmodes. Indeed if $r= v_{\min}(A^\dagger A)$ holds,
where $v_{\min}$ is the eigenvector of the smallest eigenvalue,
then it follows
\begin{equation}
\norm{e} = \norm{A^{-1} r} = \norm{A^{-1}} \norm{r} = \frac{\norm{r}}{\sqrt{\lambda_{\min}(A^\dagger A)}}~.
\end{equation}

\section{Proof of concept for the initial guesses} \label{sec:proof}

As discussed in Section \ref{sec:init_guess}, we generate
an initial guess for solving the $(n+1)$th shifted linear system
based on Lagrangian interpolation of the previous $n$ solutions.
The Lagrangian interpolation of a function $f(x)$ is given by
\begin{equation}
	L_k(x) = \sum_{i=1}^{k}f(x_i)l_{i,k}(x)
\end{equation}
where $k>1$ and $l_{i,k}(x)$ satisfies $l_{i,k}(x_j) = \delta_{ij}$ for all $i,j\in[1,k]$ with $\delta_{ij}$ being the Kronecker delta.
A polynomial solution to the latter property is
\begin{equation}
l_{i,k}(x) = \prod_{\substack{j=1\\j\neq i}}^{k} \frac{x - x_j}{x_i-x_j} \quad\Longrightarrow\quad l_{i,k}(x_j) = \delta_{ij} \text{ for } i,j\in[1,k]
\end{equation}
where the Lagrangian interpolation defined via $l_{i,k}(x)$ is denoted as the Lagrange's form.
In this case $L_k(x)$ is the unique polynomial of degree $(k-1)$ which exactly interpolates $k$ fixed points of the function $f(x)$,
i.e. $L(x_i)=f(x_i)$ for all $i\in[1,k]$.
Additionally we define $l_{1,1} (x) \equiv 1$ which gives
\begin{equation}\label{eq:lagrange_id}
\sum_{i=1}^{k} l_{i,k}(x) = 1\qquad\text{for all }k\in \mathbb{N}^+~.
\end{equation}

Furthermore it follows that the Lagrangian interpolation of a constant is exact, 
$L_k(x) = \sum_{i=1}^{k}b\,l_{i,k}(x) = b \sum_{i=1}^{k}l_{i,k}(x)\equiv b$.
The Lagrangian interpolation of $(Q+mI)^{-1}$ with grid points $\{(Q+m_i I)^{-1} \}$ is given by
\begin{equation}
L_k(m) = \sum_{i=1}^{k} l_{i,k}(m) (Q+m_iI)^{-1} \quad\text{with }\ l_{i,k}(m) = \prod_{\substack{j=1\\j\neq i}}^{k} \frac{m - m_j}{m_i-m_j}.
\end{equation}
The interpolated solution reads as
\begin{equation}\label{eq:solutions_interpolation}
\tilde x_k(m)=L_k(m)b = \sum_{i=1}^{k} l_{i,k}(m) (Q+m_iI)^{-1}b = \sum_{i=1}^{k} l_{i,k}(m) x_i
\end{equation}
where $x_i$ are solutions of $(Q+m_iI) x_i = b$
computed with a residual $r_i = b-(Q+m_iI) x_i$ which fulfills the solver stopping criterium $\norm{r_i}<\rho\norm{b}$.
For the residual of the interpolated solution $\tilde x_k(m)$ follows
\begin{align}\label{eq:residual_interpolation}
\tilde r_k(m)&=b-(Q+mI)\tilde x_k(m) = b- \sum_{i=1}^{k} l_{i,k}(m) (Q+mI)x_i \\
&= \sum_{i=1}^{k} l_{i,k}(m) \left((m_i-m)x_i + b-(Q+m_iI)x_i\right) =\sum_{i=1}^{k} l_{i,k}(m) \left((m_i-m)x_i + r_i\right)\nonumber
\end{align}
which is a Lagrangian interpolation of the residuals, i.e.~$\tilde r_k(m_i) = r_i$.
Studying the norm of the Lagrange's form of the residuals we obtain
\begin{equation}
\norm{\sum_{i=1}^{k} l_{i,k}(m) r_i} \leq \sum_{i=1}^{k} \abs{l_{i,k}(m)} \norm{r_i} < \rho \norm{b} \sum_{i=1}^{k} \abs{l_{i,k}(m)}  = \rho\,\gamma_{k}(m) \norm{b}
\end{equation}
where $\gamma_{k}(m) = \sum_{i=1}^{k} \abs{l_{i,k}(m)}$ is the Lebesgue function defined from the Lagrange's polynomials $l_{i,k}(m)$.
The Lebesgue function in the interval $[m_{\min},m_{\max}]$ assumes values
\begin{equation}
1 \leq \gamma_{k}(m) \leq \Gamma_k
\end{equation}
where $m_{\min}$ and $m_{\max}$ are the smallest and largest shifts of $m_i$, respectively, 
while $\Gamma_k= \max_{m\in [m_{\min},m_{\max}]}\gamma_{k}(m)$ is referred 
to as Lebesgue constant. Depending on the shifts, the Lebesgue constant could grow as an exponential,
logarithmic or asymptotic function of $k$~\cite{smith2006lebesgue}.
In case of diverging growths one can truncate the degree of the interpolation in order to keep the error under control.
The relation 
\begin{equation}
\norm{\sum_{i=1}^{k} l_{i,k}(m) r_i} \leq  \rho\,\Gamma_k \norm{b}
\end{equation}
fixes the maximal contribution of the residuals to the residual of the interpolated solution.

The additional term to the Lagrange's form in Eq.~(\ref{eq:residual_interpolation}) can be re-written as
\begin{align}
\sum_{i=1}^{k} l_{i,k}(m) (m-m_i)x_i & \simeq \left(\prod_{j=1}^{k} (m - m_j)\right) \sum_{i=1}^{k} \left(\prod_{\substack{j=1\\j\neq i}}^{k}\frac{1}{m_i-m_j}\right) (Q+m_iI)^{-1} b \nonumber \\
&= \left(\prod_{j=1}^{k} \frac{m - m_j}{Q+m_jI}\right) b
\end{align}
where $x_i\simeq (Q+m_i I)^{-1} b $ is used and the partial fraction decomposition is re-summed in a product of fractions.
For the norm of follows
\begin{equation}
\norm{\sum_{i=1}^{k} l_{i,k}(m) (m-m_i)x_i} \simeq  \norm{\left(\prod_{i=1}^{k} \frac{m - m_i}{Q+m_iI}\right) b} \leq \prod_{i=1}^{k} \frac{\abs{m - m_i}}{\abs{\lambda_{\min,i}}} \norm{b}
\end{equation}
where $\lambda^2_{\min,i}$ is the smallest eigenvalue of $(Q+m_i)^\dagger(Q+m_i)$.
If $Q$ is a positive-definite matrix then $\lambda_{\min,i} = \lambda_{\min} +m_i$ with $\lambda_{\min}>0$ being the smallest eigenvalue of $Q$. 

Considering now the full residual interpolation in Eq.~(\ref{eq:residual_interpolation}) we find the following upper limits
\begin{align}\label{eq:residual_interpolation_lim}
\norm{\tilde r_k(m)} &= \norm{\sum_{i=1}^{k} l_{i,k}(m) \left((m-m_i)x_i + r_i\right)} \leq \norm{\sum_{i=1}^{k} l_{i,k}(m) (m-m_i)x_i }  + \rho\,\gamma_k(m)\norm{b} \nonumber\\
&\lesssim \left(\prod_{i=1}^{k} \frac{\abs{m - m_i}}{\abs{\lambda_{\min,i}}} + \rho\,\Gamma_k\right) \norm{b}.
\end{align}
The first upper bound requires a knowledge of the solutions $x_i$ and residuals $r_i$, making it dependent on the numerical approach. The second instead depends only on the analytical properties of the shifted systems.

\end{document}